\documentclass[twocolumn,superscriptaddress,prd]{revtex4}
\usepackage{graphicx}
\usepackage{subfigure}
\usepackage{dcolumn}
\usepackage{bm}
\begin{document}
\topmargin 0.0001cm
\title{Search for  the $\Theta^+$ pentaquark in the reactions
$\gamma p \to \bar K^0K^+n$ and $\gamma p \to \bar K^0K^0p$\\}


\newcommand*{\INFNGE}{Istituto Nazionale di Fisica Nucleare, Sezione di Genova, and Dipartimento di Fisica, Universit\`a di Genova,
 16146 Genova, Italy}
\affiliation{\INFNGE}
\newcommand*{\RPI}{Rensselaer Polytechnic Institute, Troy, New York 12180-3590}
\affiliation{\RPI}
\newcommand*{\SCAROLINA}{University of South Carolina, Columbia, South Carolina 29208}
\affiliation{\SCAROLINA}
\newcommand*{\CMU}{Carnegie Mellon University, Pittsburgh, Pennsylvania 15213}
\affiliation{\CMU}
\newcommand*{\JLAB}{Thomas Jefferson National Accelerator Facility, Newport News, Virginia 23606}
\affiliation{\JLAB}
\newcommand*{\RICE}{Rice University, Houston, Texas 77005-1892}
\affiliation{\RICE}
\newcommand*{\UCONN}{University of Connecticut, Storrs, Connecticut 06269}
\affiliation{\UCONN}

\newcommand*{\ANL}{Argonne National Laboratory}
\affiliation{\ANL}
\newcommand*{\ASU}{Arizona State University, Tempe, Arizona 85287-1504}
\affiliation{\ASU}
\newcommand*{\UCLA}{University of California at Los Angeles, Los Angeles, California  90095-1547}
\affiliation{\UCLA}
\newcommand*{\CSU}{California State University, Dominguez Hills, California  90747-0005}
\affiliation{\CSU}
\newcommand*{\CUA}{Catholic University of America, Washington, D.C. 20064}
\affiliation{\CUA}
\newcommand*{\SACLAY}{CEA-Saclay, Service de Physique Nucl\'eaire, F91191 Gif-sur-Yvette, France}
\affiliation{\SACLAY}
\newcommand*{\CNU}{Christopher Newport University, Newport News, Virginia 23606}
\affiliation{\CNU}
\newcommand*{\ECOSSEE}{Edinburgh University, Edinburgh EH9 3JZ, United Kingdom}
\affiliation{\ECOSSEE}
\newcommand*{\FU}{Fairfield University, Fairfield CT 06824}
\affiliation{\FU}
\newcommand*{\FIU}{Florida International University, Miami, Florida 33199}
\affiliation{\FIU}
\newcommand*{\FSU}{Florida State University, Tallahassee, Florida 32306}
\affiliation{\FSU}
\newcommand*{\GWU}{The George Washington University, Washington, DC 20052}
\affiliation{\GWU}
\newcommand*{\ECOSSEG}{University of Glasgow, Glasgow G12 8QQ, United Kingdom}
\affiliation{\ECOSSEG}
\newcommand*{\ISU}{Idaho State University, Pocatello, Idaho 83209}
\affiliation{\ISU}
\newcommand*{\INFNFR}{INFN, Laboratori Nazionali di Frascati, 00044 Frascati, Italy}
\affiliation{\INFNFR}
\newcommand*{\ORSAY}{Institut de Physique Nucleaire ORSAY, Orsay, France}
\affiliation{\ORSAY}
\newcommand*{\IHEP}{Institute for High Energy Physics, Protvino, 142281, Russia}
\affiliation{\IHEP}
\newcommand*{\ITEP}{Institute of Theoretical and Experimental Physics, Moscow, 117259, Russia}
\affiliation{\ITEP}
\newcommand*{\JMU}{James Madison University, Harrisonburg, Virginia 22807}
\affiliation{\JMU}
\newcommand*{\KHARKOV}{Kharkov Institute of Physics and Technology, Kharkov 61108, Ukraine}
\affiliation{\KHARKOV}
\newcommand*{\KYUNGPOOK}{Kyungpook National University, Daegu 702-701, South Korea}
\affiliation{\KYUNGPOOK}
\newcommand*{\UMASS}{University of Massachusetts, Amherst, Massachusetts  01003}
\affiliation{\UMASS}
\newcommand*{\MOSCOW}{Moscow State University, General Nuclear Physics Institute, 119899 Moscow, Russia}
\affiliation{\MOSCOW}
\newcommand*{\UNH}{University of New Hampshire, Durham, New Hampshire 03824-3568}
\affiliation{\UNH}
\newcommand*{\NSU}{Norfolk State University, Norfolk, Virginia 23504}
\affiliation{\NSU}
\newcommand*{\OHIOU}{Ohio University, Athens, Ohio  45701}
\affiliation{\OHIOU}
\newcommand*{\ODU}{Old Dominion University, Norfolk, Virginia 23529}
\affiliation{\ODU}
\newcommand*{\URICH}{University of Richmond, Richmond, Virginia 23173}
\affiliation{\URICH}
\newcommand*{\UNIONC}{Union College, Schenectady, NY 12308}
\affiliation{\UNIONC}
\newcommand*{\VT}{Virginia Polytechnic Institute and State University, Blacksburg, Virginia   24061-0435}
\affiliation{\VT}
\newcommand*{\VIRGINIA}{University of Virginia, Charlottesville, Virginia 22901}
\affiliation{\VIRGINIA}
\newcommand*{\WM}{College of William and Mary, Williamsburg, Virginia 23187-8795}
\affiliation{\WM}
\newcommand*{\YEREVAN}{Yerevan Physics Institute, 375036 Yerevan, Armenia}
\affiliation{\YEREVAN}
\newcommand*{\UK}{University of Kentucky, Lexington, Kentucky 40506}
\affiliation{\UK}
\newcommand*{\UNCW}{University of North Carolina, Wilmington, North Carolina 28403}
\affiliation{\UNCW}
\newcommand*{\UAT}{North Carolina Agricultural and Technical State University, Greensboro, North Carolina 27455}
\affiliation{\UAT}
\newcommand*{\RIKEN}{The Institute of Physical and Chemical Research, RIKEN, Wako, Saitama 351-0198, Japan}
\affiliation{\RIKEN}
\newcommand*{\NOWUNH}{University of New Hampshire, Durham, New Hampshire 03824-3568}
\newcommand*{\NOWUMASS}{University of Massachusetts, Amherst, Massachusetts  01003}
\newcommand*{\NOWMIT}{Massachusetts Institute of Technology, Cambridge, Massachusetts  02139-4307}
\newcommand*{\NOWODU}{Old Dominion University, Norfolk, Virginia 23529}
\newcommand*{\NOWSCAROLINA}{University of South Carolina, Columbia, South Carolina 29208}
\newcommand*{\NOWGEISSEN}{Physikalisches Institut der Universit\"at Gie{\ss}en, 35392 Giessen, Germany}
\newcommand*{\NOWECOSSEE}{Edinburgh University, Edinburgh EH9 3JZ, United Kingdom}
\newcommand*{\NOWNONE}{unknown, }

\author {R.~De~Vita} 
\affiliation{\INFNGE}
\author {M.~Battaglieri} 
\affiliation{\INFNGE}
\author {V.~Kubarovsky} 
\affiliation{\RPI}
\author {N.A.~Baltzell} 
\affiliation{\SCAROLINA}
\author {M.~Bellis} 
\affiliation{\RPI}
\affiliation{\CMU}
\author {J.~Goett} 
\affiliation{\RPI}
\author {L.~Guo} 
\affiliation{\JLAB}
\author {G.S.~Mutchler} 
\affiliation{\RICE}
\author {P.~Stoler} 
\affiliation{\RPI}
\author {M.~Ungaro} 
\affiliation{\RPI}
\affiliation{\UCONN}
% ...

\collaboration{The CLAS Collaboration}
     \noaffiliation
%
 
%
%


\date{\today}

\begin{abstract}
The exclusive reactions $\gamma p \to \bar K^0 K^+ n$ and $\gamma p \to \bar K^0 K^0 p$ have been studied 
in the photon energy range 1.6--3.8 GeV, searching for  evidence 
of the exotic baryon  $\Theta^+(1540)$ in the decays $\Theta^+\to nK^+$ and $\Theta^+\to p K^0$.
Data were collected with the 
CLAS detector at the Thomas Jefferson National Accelerator Facility.  The integrated luminosity
was about  70 pb$^{-1}$. The  reactions have been isolated by detecting the $K^+$ and proton directly,
the neutral kaon  via its decay to  $K_S \to \pi^+ \pi^-$ and the neutron or neutral kaon via the missing mass technique.
The mass and width of known hyperons such as $\Sigma^+$,  $\Sigma^-$ and $\Lambda(1116)$
were used as a check of the mass determination accuracy and experimental resolution.
Approximately 100,000 $\Lambda^*(1520)$'s  and 150,000 $\phi$'s were observed in the $\bar K^0  K^+ n$ and
$\bar K^0  K^0 p$ final state respectively.
No evidence for the $\Theta^+$ pentaquark was found in the $nK^+$ or $pK_S$ invariant mass spectra.
Upper limits were set on the  production cross section of the reaction $\gamma p \to \Theta^+ \bar K^0$ 
as functions of center-of-mass angle,  $nK^+$ and $pK_S$ masses.
Combining the results of the two reactions, the 95\% C.L.  upper limit on the total cross section for a resonance 
peaked at 1540 MeV was found to be 0.7 nb. Within  most of the available theoretical models, 
this corresponds to an upper limit on the  $\Theta^+$ width,  $\Gamma_{\Theta^{+}}$, ranging between 0.01  and 7 MeV.

\end{abstract}

\maketitle

\section{\label{sec:intro}Introduction}
The possible existence of baryon states beyond the usual $qqq$ 
configuration is of fundamental importance for the 
understanding of hadronic structure. 
QCD does not prohibit the existence of exotic
states with different configurations such as
$qqqq\bar q$. In fact, measurements of nucleon structure 
functions from high energy lepton-nucleon experiments have shown
for example that ``sea'' quarks ($q\bar q$ pairs) contribute 
significantly to the total momentum
of the nucleon.
Indeed  usual baryons are  admixtures of the standard $qqq$ configuration and of 
$qqqq\bar q$, $qqqg$, etc.

In the past, experimental 
searches focused on  pentaquarks, 
{\it i.e.} baryons with a minimal $qqqq \bar q$ structure.
In 1997,  Diakonov and collaborators~\cite{diakonov} made definite 
predictions about the masses and widths of a decuplet of pentaquark 
states (the so-called ``antidecuplet'') in the framework of a chiral soliton model. 
The most intriguing 
aspect of such a multiplet is the presence of three states 
with exotic quantum numbers or a combination of quantum numbers not allowed for ordinary baryons:
 the $\Theta^+$ with  $S$=+1,  the 
$\Xi^{--}$ and $\Xi^{+}$ with $S=-2$. In particular the 
 positive  strangeness $\Theta^+$ is not compatible with a $qqq$ state, requiring a minimal 
quark configuration of  the  type $uudd\bar{s}$. 
The widths of the exotic pentaquarks were predicted by this 
model to be very narrow (10--15 MeV) implying that if 
such states exist they should be directly visible in
invariant mass spectra  without the need for a more sophisticated partial 
wave analysis.
\begin{figure}
\vspace{8.2cm} 
\includegraphics{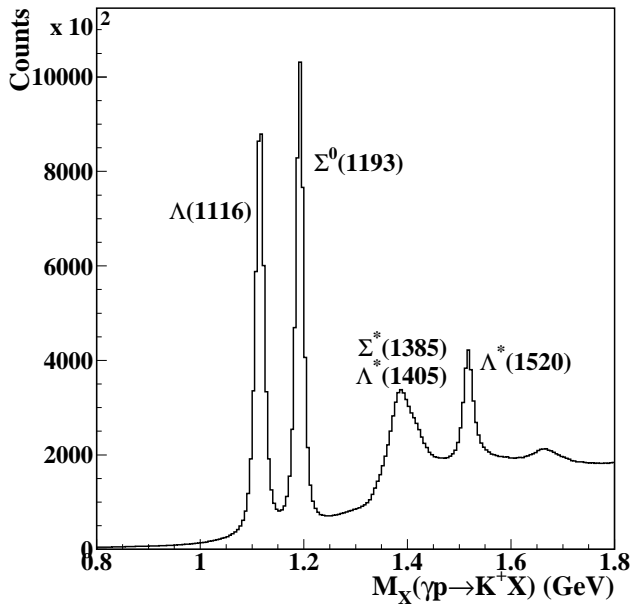}
\caption[]{The $K^+$ missing mass spectrum for the reaction  $\gamma p \to  K^+ X$. Peaks correspond to ground and 
excited states of well-known hyperons.}
\label{fig:k_spec}
\end{figure}

The first evidence of a  $\Theta^+$ candidate
was reported in October 2002 by the LEPS Collaboration, 
based on the re-analysis of existing data~\cite{nakano}.
Several other experimental groups followed~\cite{itep,clas,saphir,clas2,itep2,hermes,zeus,serpu,cosytof,jinr}
reporting evidence of a peak in the
mass range 1521--1555 MeV. 
The observation of an isospin partner of the $\Theta^+$ (the $\Theta^{++}$) was recently reported by the 
STAR Collaboration~\cite{star} while 
the observation of a second pentaquark state,  the $\Xi^{--}$ with $dsds\bar{u}$ structure, 
was reported by the NA49 Collaboration \cite{na49}. The  first evidence for an
anti-charmed pentaquark, $\Theta_c$, was reported  by the H1 Collaboration \cite{h1}. 
While pentaquark signals observed in each experiment suffered from low statistics,
the observations in many different reactions using different probes (photons, electrons, protons, neutrinos)
and targets (protons, neutrons, nuclei) supported the pentaquark's existence. 
On the contrary, subsequent re-analysis of data collected in a different set of experiments
\cite{aleph,babar,belle,bes,cdf,focus,herab,hypercp,lass,l3,phenix,sphinx,wa89,compass}  found
no evidence of pentaquarks casting doubts about their existence.

The experimental evidences, both positive and negative, were
obtained from data previously collected for other purposes in  
many reaction channels and in diverse  kinematic conditions, 
thus may involve different production mechanisms. 
As a result, direct comparisons of  the different experiments 
are very difficult,  preventing a definitive conclusion about the pentaquark's existence. 
A second generation of dedicated experiments~\cite{prl,g10}, optimized for the 
pentaquark search, was undertaken at the Department of Energy's  
Thomas Jefferson National Accelerator Facility. 
These experiments covered the few GeV region in photon energy, 
 where most of the positive evidence was reported,
and collected  at least an order of magnitude more data than used in the previous measurements.

The exclusive reactions $\gamma p \to \bar K^0 K^+ n$ and $\gamma p \to \bar K^0 K^0 p$
were studied with the CLAS detector~\cite{B00} with 1.6 to 3.8 GeV energy photons, to look for evidence
of the reaction $\gamma p \to \bar K^0 \Theta^+$, where the  $\Theta^+$ decays into $K^0 p$ or $K^+ n$.
The main results for the $\gamma p \to \bar K^0 K^+ n$ channel were reported in Ref.~\cite{prl}.
In this paper we discuss in detail the analysis procedure and the results for both decay modes, 
combining them to give a final consistent result.

For the  $\Theta^+ \to n K^+$ decay mode, the measurement of all participating particles allows one to tag the strangeness 
of the reaction which clearly identifies the exotic nature of the baryon  produced 
in association with the  $\bar K^0$.
For the other possible decay mode,  $\Theta^+ \to p K^0$, since  a $K_S$ was measured, the strangeness of the 
$p K_S$ invariant mass system is not defined. Nevertheless, we were motivated to analyze this channel since 
the majority of the positive 
results~\cite{itep,hermes,zeus,serpu,cosytof,jinr} were reported looking at this decay mode. 
Moreover the CLAS acceptance for the two reactions is
somewhat complementary and the combination of the two channels  results in  complete kinematic coverage.
According to many theoretical predictions, e.g.~\cite{theory-treshprod,oh04,Guidal,Nam}, 
the photon energy range covered by  this experiment should be the best to explore 
since the $\Theta^+$ is expected to have its maximum  cross section 
near the production threshold. Also, in this kinematic region,
the CLAS detector provides a mass resolution of few MeV and an accuracy in the
mass determination of 1--2 MeV, which is  necessary 
to pin down the mass and width of any narrow peak in the spectrum.

The $\gamma p \to \bar K^0 K^+ n$ channel was previously investigated  at ELSA by the SAPHIR collaboration~\cite{saphir} 
in a similar photon energy range, finding positive evidence for a narrow 
$\Theta^+$ state with $M=1540$ MeV and FWHM $\Gamma$ less than 25 MeV.
The most recent analysis resulted in a total production cross section of the order of 
50 nb.
Since this experiment completely overlaps the kinematic regions of the SAPHIR experiment, 
the new results put those previous findings to a direct test for the first time.
Results on pentaquark searches in the exclusive 
reaction  $\gamma p \to \bar K^0 K^0 p$ have never been published before.

In the following, some details are given on the experiment (Sec.~\ref{sec:exp})
 and its analysis (Sec.~\ref{sec:anal}). 
The findings of  $\gamma p \to \bar K^0 K^+ n$ channel  are compared with the SAPHIR
experiment in Sec.~\ref{sec:comp} and  the  systematic checks are discussed in Sec.~\ref{sec:syst}. 
The final results are reported in the last section.

\begin{figure}
\vspace{7.5cm} 
\includegraphics{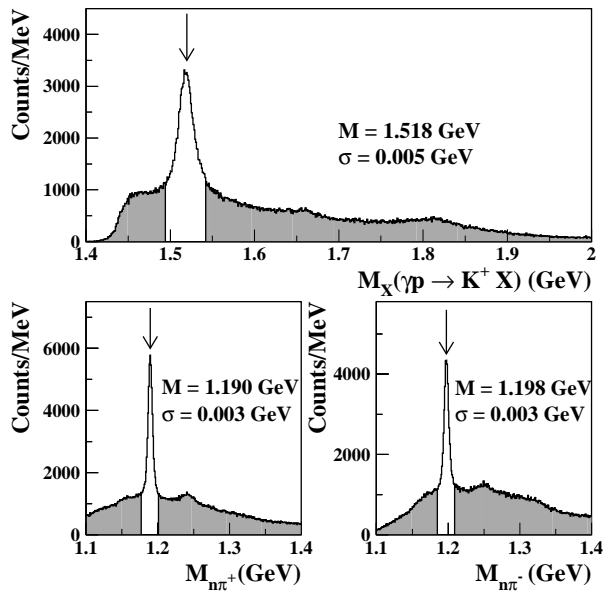}
\caption[]{Hyperon production in the $\gamma p \to   \bar K^0 K^+ n$ reaction.
 Top: $K^+$ missing mass spectrum with the $\Lambda^*(1520)$. 
Bottom: $n\pi^+$ (left) and $n\pi^-$ (right) invariant mass spectra. 
The mass position and width of the measured peaks are shown (for the $\Lambda^*$  the experimental resolution is reported).
 For comparison the arrows show the world data value for the mass position. The shaded 
area indicates the retained events.}
\label{fig:hyper}
\end{figure}
\begin{figure}
\vspace{7.5cm} 
\includegraphics{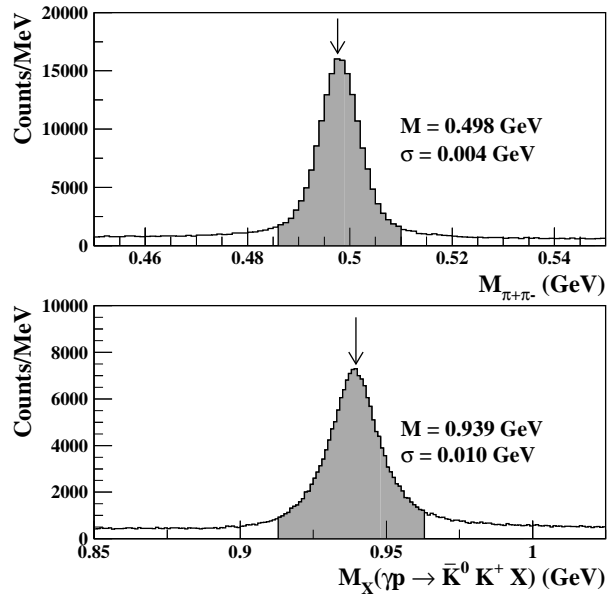}
\caption[]{Mass plots for analysis of the reaction  $\gamma p \to  \bar K^0 K^+ n$.
Top: $\pi^+ \pi^-$ invariant mass and the $\bar K^0$ peak. 
Bottom:  missing mass for the reaction $\gamma p \to  \bar K^0 K^+ X$ and the neutron peak. 
The labels show the mass position and width of the measured peaks. 
For comparison the arrows show the world averages for the mass positions. The shaded area indicates the retained events.}
\label{fig:pid}
\end{figure}

\section{\label{sec:exp} The experiment}
The measurement   was performed using the CLAS detector 
in  Hall B at Jefferson Lab with a bremsstrahlung 
photon beam  produced by a continuous 60 nA electron beam of  $E_0$ = 4.02 GeV  
impinging on a gold foil of thickness $8 \times 10^{-5}$ radiation lengths.
A bremsstrahlung tagging system~\cite{SO99} with a photon energy resolution of 0.1$\%$ $E_0$ 
was used to tag photons in the energy range from 1.58 GeV (about the $\Theta^+(1540)$ production threshold)
to a maximum energy of 3.8 GeV.
A cylindrical liquid hydrogen target cell 4 cm in diameter and 40 cm long was used.
Outgoing  hadrons were detected in the CLAS  spectrometer.
Momentum information for charged particles was obtained via tracking
through three regions of multi-wire drift chambers~\cite{DC} in conjunction with  a toroidal magnetic 
field ($\sim 0.5$ T) generated by six superconducting coils. 
The polarity of the field was set to bend the positive particles away from the beam into the acceptance 
of the  detector.
Time-of-flight scintillators (TOF) were used for charged hadron
identification~\cite{Sm99}. 
The interaction time between the incoming photon and the target
was measured by the start counter (ST)~\cite{ST}. This   is
made of 24 strips of 2.2 mm thick plastic scintillator surrounding the hydrogen cell
with a single-ended PMT-based read-out. A time resolution of $\sim$300 ps  was achieved. 
The CLAS momentum resolution, $\sigma_p/p$ is from 0.5 to 1\%, depending on
the kinematics.  CLAS is well suited for simultaneous multi-hadron 
detection as required by  experiments searching for pentaquarks 
(this experiment required at least 3 hadrons detected). 
The detector geometrical acceptance for each positive particle in the 
relevant kinematic region is about 40\%. It is somewhat less for low energy negative 
hadrons, which can be lost at  forward angles because
their paths are bent toward the beam line and out of the acceptance
by the toroidal field.
Coincidences between the photon tagger and the CLAS detector triggered 
the recording of the events. The trigger in CLAS  required
a coincidence between the TOF and the ST 
in at least two sectors, in order to  select
reactions with at least two charged particles in the final state.
The collected data sample contains
events from several reaction  channels in addition to the reactions of interest.
Reactions such as  
$\gamma p \to p \pi^+ \pi^-$, $\gamma p \to p \omega$, 
$\gamma p \to   K^+ X$, and $\gamma p \to  \Sigma^{+(-)} \pi^{-(+)}  K^+$
have been used for systematic studies.
An integrated luminosity of 70 pb$^{-1}$ 
was accumulated in  50 days of running  in 2004.

\section{Data Analysis}\label{sec:anal}
The raw data were passed through the standard CLAS reconstruction software to determine the 4-momenta of detected particles.
In this phase of the analysis, corrections were applied to account for the energy loss of charged particles in the target and 
surrounding  materials, misalignments of the  drift chambers' position, and 
uncertainties in the value of the toroidal magnetic field.
The  energy calibration
of the Hall-B tagger system
was performed both by a direct measurement of the $e^+e^-$ pairs produced by the incoming photons~\cite{tag-abs_cal}
and by applying  an over-constrained kinematic fit  to the  reaction $\gamma p \to p \pi^+ \pi^-$, where all particles
in the final state were detected in CLAS~\cite{tag-kinefit}.
The quality of the calibrations was checked by 
looking at  the mass  of known particles as well as their dependence on  other kinematic variables 
(photon energy, detected particle momenta and angles).
As an example, Fig.~\ref{fig:k_spec} shows the $K^+$ missing
mass spectrum of the reaction $\gamma p \to K^+  X$:
peaks of known hyperons such as the $\Lambda(1116)$, the $\Sigma^0(1193)$, and related excited states are clearly visible.

\subsection{Pentaquark Analysis Strategy}
The  data set was independently analyzed by three groups that made use of different analysis tools and procedures. 
This strategy was adopted both to have a corroboration of the analysis results
 and an estimate of the systematic errors associated with the choice of the analysis procedure. 
All three analyses agreed on the main conclusions. 
In the following sections we report in detail the analysis procedure of one  group
 while the comparison of the results from the different groups is discussed in Section~\ref{sec:ind}.
\begin{figure}
\vspace{8.0cm} 
\includegraphics{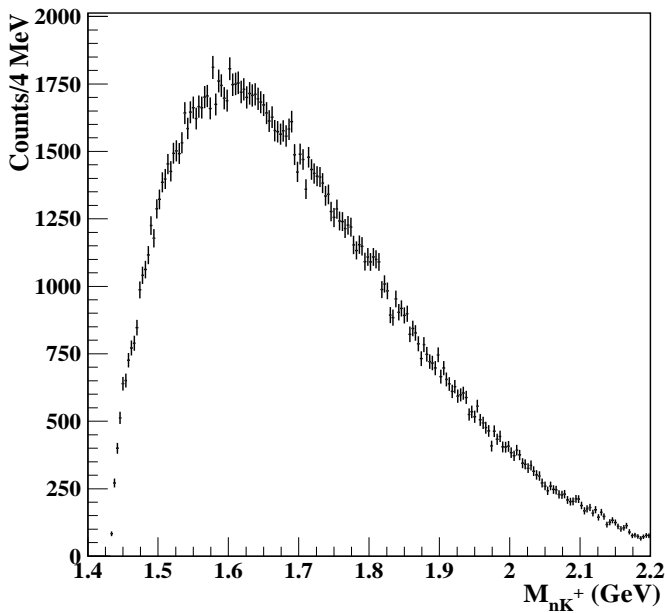}
\caption[]{The $nK^+$ invariant mass spectrum for the reaction $\gamma p \to  \bar K^0 K^+ n$ after all cuts,  obtained by integrating over all measured photon energies and $\bar K^0$ angles.}
\label{fig:nkp_spectrum}
\end{figure}

\subsection{Reaction identification: $\gamma p \to \bar K^0K^+n$}\label{ssec:reac1}
The reaction $\gamma p \to \bar K^0 K^+ n$ was isolated as follows.
The $K^+$ was  directly detected by the spectrometer, while the $K_S$ component of the $\bar K^0$ 
was reconstructed from the invariant  mass of its  $\pi^+\pi^-$ decay (B.R. $\sim$ 69\%). 
The neutron was then reconstructed from the missing mass of all the detected particles.
After its identification, the neutron mass used to calculate other kinematic variables was 
kept fixed at its  nominal PDG value~\cite{PDG}. A $\pm 3 \sigma$ cut around the $K_S$ and the neutron peaks 
was applied to isolate the exclusive reaction.
A total of 320,000 events was selected by this procedure. 
Three background reactions having the same final state as the reaction of 
 interest were clearly identified: $\gamma p  \to  K^+ \Lambda^*(1520)$
with $\Lambda^*(1520) \to n \bar K^0$, 
$\gamma p  \rightarrow  K^+ \Sigma^+ \pi^-$, and $\gamma p  \rightarrow  K^+ \Sigma^- \pi^+$ with
$\Sigma^{+(-)} \to n \pi^{+(-)}$. 
Figure~\ref{fig:hyper} shows the background hyperon peaks: the
$\Lambda^*(1520)$ in the  $K^+$ missing mass spectrum  and
the  $\Sigma^+$ and   $\Sigma^-$ peaks in the $n\pi^+$  and $n\pi^-$
invariant mass spectra respectively.  We found  $M_{\Sigma^+} = 1190 \pm 1$ MeV and 
$M_{\Sigma^-} =1198 \pm1$ MeV,  with a measured experimental width $\sigma \sim  $3.5 MeV for both of them.
These  are in excellent agreement with the world data~\cite{PDG},
and are a measure of the quality of the mass determination. Since  these states have a much smaller width than  the CLAS
resolution, their observed widths provide an estimate of the 
experimental resolution.  The reported values are in good  agreement with the
CLAS resolution estimated from simulations.
To remove the  contribution of these  channels,
a $\pm 3 \sigma$ cut  was applied around the $\Sigma$  peaks while a $\pm$24 MeV cut was applied 
around the $\Lambda^*(1520)$ peak, resulting in  a total of 160,000  retained events.
The resulting  $K_S$ and neutron mass plots are shown in Fig.~\ref{fig:pid}, 
where the two peaks are seen above
 small background levels. 
The peak positions of the  reconstructed $K_S$ and neutron masses 
were found to be 498$\pm$1 MeV and 939$\pm$1 MeV respectively.
\begin{figure}[h]
\vspace{8.5cm} 
\includegraphics{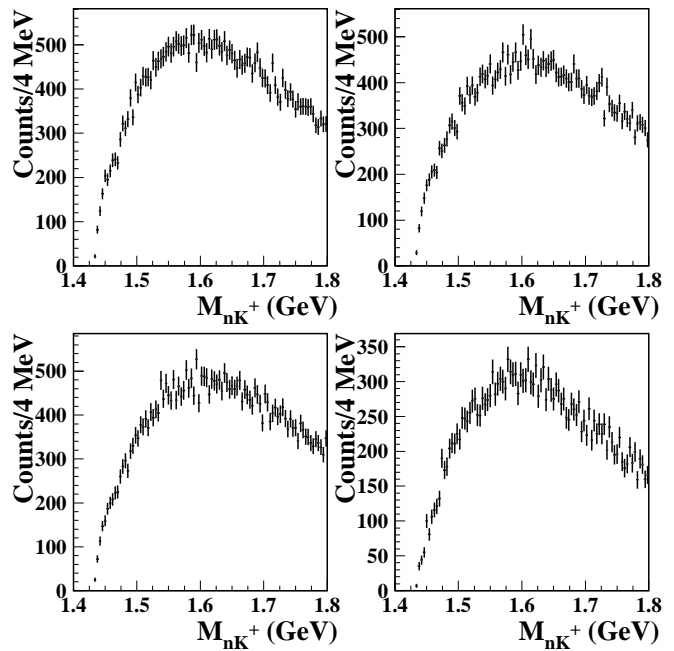}
\caption[]{The $nK^+$ invariant mass spectrum for the reaction $\gamma p \to  \bar K^0 K^+ n$ for different $\cos\theta^{CM}_{\bar K^0}$ ranges: 
$\cos\theta^{CM}_{\bar K^0}<-0.5$ (top left), $-0.5<\cos\theta^{CM}_{\bar K^0}<0$ (top right), 
$0<\cos\theta^{CM}_{\bar K^0}<0.5$ (bottom left), $\cos\theta^{CM}_{\bar K^0}>0.5$ (bottom right).}
\label{fig:nkp_angcut}
\end{figure}

After all cuts, the resulting $nK^+$ invariant mass spectrum is shown 
in Fig.~\ref{fig:nkp_spectrum}. The spectrum is smooth and structureless. In particular no evidence for 
a peak or an  enhancement is observed at masses close to 1540 MeV, where signals 
associated with the $\Theta^+$ were previously reported. 
To enhance a possible resonance signal not visible in the integrated spectra, 
we assumed  the  two-body reaction $\gamma p \to \bar K^0 \Theta^+(1540)$
and selected different $\bar K^0$ center-of-mass angle intervals.
Fig.~\ref{fig:nkp_angcut} shows the $nK^+$ invariant mass spectrum for different
$\cos\theta^{CM}_{\bar K^0}$ ranges.
Monte Carlo studies of the CLAS acceptance for this reaction (see Sec.~\ref{ssec:ul}) showed that 
we could detect  events over the entire angular range,
in spite of  a drop in the  efficiency at forward $\bar K^0$ angles,
$\theta^{CM}_{\bar K^0}<30^\circ$.
No structures were found in the spectra when specific angular ranges were selected.

\begin{figure}
\vspace{7.5cm} 
\includegraphics{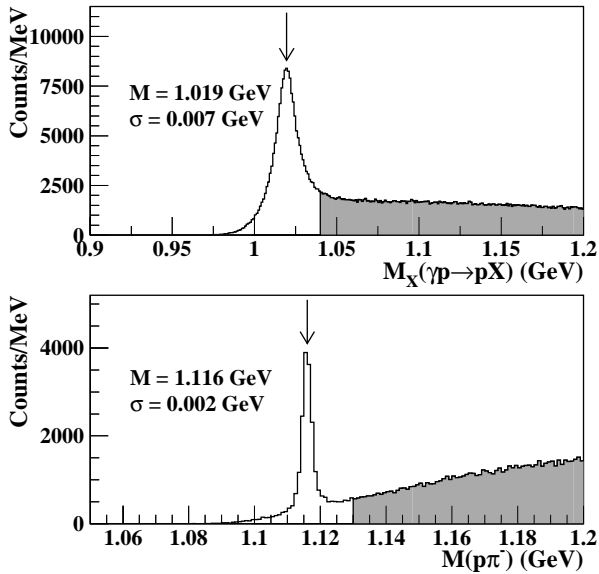}
\caption[]{Background reactions  in $\gamma p \to   \bar K^0 K^0 p$ reaction.
Top: proton missing mass spectrum with the $\phi$ peak. 
Bottom: $p \pi^-$ invariant mass spectrum; the $\Lambda(1116)$ peak is clearly visible.
The mass position and width of the measured peaks are shown. For comparison the arrows show the world data value for the mass position. The shaded 
area indicates the retained events.}
\label{fig:hyper2}
\end{figure}

\begin{figure}
\vspace{7.5cm} 
\includegraphics{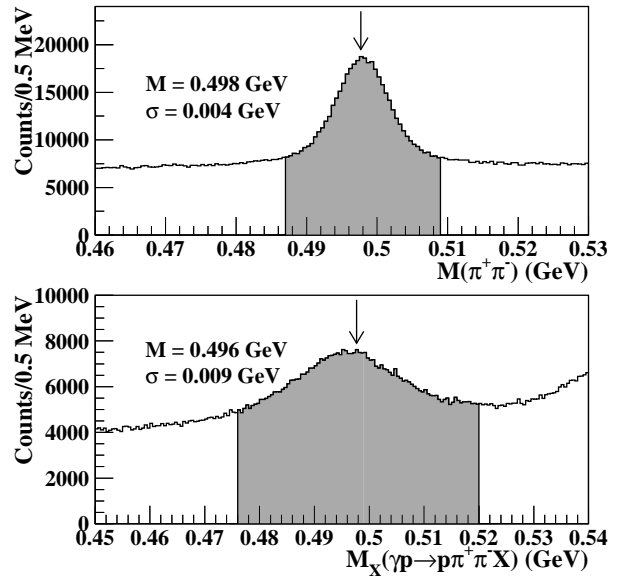}
\caption[]{Mass plots for analysis  of  the reaction $\gamma p \to \bar K^0 K^0 p$. 
Top: $\pi^+ \pi^-$ invariant mass and the $K_S$ peak. 
Bottom:  missing mass for the reaction $\gamma p \to   K_S p X$ and the neutral kaon peak.
The labels show the mass position and width of the measured peaks. 
For comparison the arrows show the world averages for the mass positions. The shaded area indicates the retained events.}
\label{fig:pid2}
\end{figure}

As a demonstration of our sensitivity to baryon resonances,
we  derived the $\Lambda^*$ yield.
We fit the  $K^+$ missing mass spectrum  of Fig.~\ref{fig:hyper} 
 by a Breit-Wigner function convoluted 
with a Gaussian function to account for the detector resolution,
plus a second-order polynomial background.
To  derive the $\Lambda^*$ yield,
following Ref.~\cite{Jackson}, we used the  Breit-Wigner form:
\begin{equation}
BW=\frac{M_0 m \Gamma(m)}{(m^2-M_0^2)^2+M_0^2\Gamma(m)^2},
\end{equation}
where:
\begin{eqnarray*}
\Gamma(m)&=&\Gamma_0\left(\frac{Q}{Q_0}\right)^{2l+1}, 
\end{eqnarray*}
$M_0$ and $\Gamma_0$ are the resonance mass and intrinsic width, $Q$
is the $\bar K^0$ momentum in the rest frame of the $n - \bar K^0$ system, 
 $Q_0$ is the same quantity evaluated at the $\Lambda^*(1520)$ peak,
and  $l$ is the $n - \bar K^0$ relative orbital 
angular momentum ($l=2$ for the $\Lambda^*$).
In the fit $\Gamma_0$ was  fixed to 15.6 MeV~\cite{PDG} while the 
$\sigma$ of the Gaussian function was allowed to vary.
Integrating the Breit-Wigner line in the mass range 1.45--2.0 GeV we obtained
a $\Lambda^*(1520)$ yield of  $99,000\pm10,000$. The quoted error is dominated by the systematic uncertainty related 
to the shape of  the Breit-Wigner and the underlying background.
The mass position was found to be 1518 $\pm$ 2 MeV, in good agreement with  world data~\cite{PDG}, 
while the  experimental resolution was found to be $\sim$ 5 MeV, typical for CLAS~\cite{B00}.

\subsection{Reaction identification: $\gamma p \to \bar K^0 K^0 p$}\label{ssec:reac2}
To isolate the reaction $\gamma p \to \bar K^0 K^0 p$, 
the proton  was  detected by the spectrometer and a  $K_S$ meson 
was reconstructed from the invariant  mass of its  $\pi^+\pi^-$ decay (B.R. $\sim$ 69\%). 
The second neutral kaon was then reconstructed from the missing mass of all the detected particles.
A $\pm 3 \sigma$ cut around the $K_S$ and the missing kaon peaks 
was applied to isolate the exclusive reaction.
A total of 750,000 events were selected by this procedure. 
The reaction $\gamma p  \to p  \phi$ with $\phi \to K_L K_S$ has  the same final state as the reaction of 
interest. The reaction $\gamma p  \rightarrow  \Lambda(1116) \pi^+ K^0 \rightarrow  p \pi^-  \pi^+  K^0$ 
also contributes to the background.
Fig.~\ref{fig:hyper2} shows the background peaks: the
$\phi$ shows up in   the proton  missing mass spectrum  and
the $\Lambda$(1116) in the  $p\pi^-$ invariant mass spectrum.
We found  $M_{\phi}$ = 1019 $\pm$ 1 MeV and 
$M_{\Lambda} =1116 \pm1$ MeV,  with a measured experimental resolution $\sigma$ of about 7 MeV and  2 MeV respectively.
The  $\phi$ peak was
fitted  with a  Breit-Wigner ($\Gamma_0=4.2$ MeV), convoluted with a Gaussian  describing the CLAS resolution, in the same way as was done for the 
 $\Lambda^*(1520)$. 
The obtained masses are in agreement with world data~\cite{PDG}.
To remove the  contribution of these channels,
only events with a $M_X(\gamma p \to p X)>$ 1.04 GeV and $M(p\pi^-)>$ 1.13 GeV 
were retained, resulting  in a  total of 550,000 events.
The two neutral kaon mass spectra after the background rejection are shown in Fig.~\ref{fig:pid2}. 
The two masses were 
found to be respectively 498$\pm$1 MeV and 496$\pm$3 MeV.

\begin{figure}
\vspace{8.0cm} 
\includegraphics{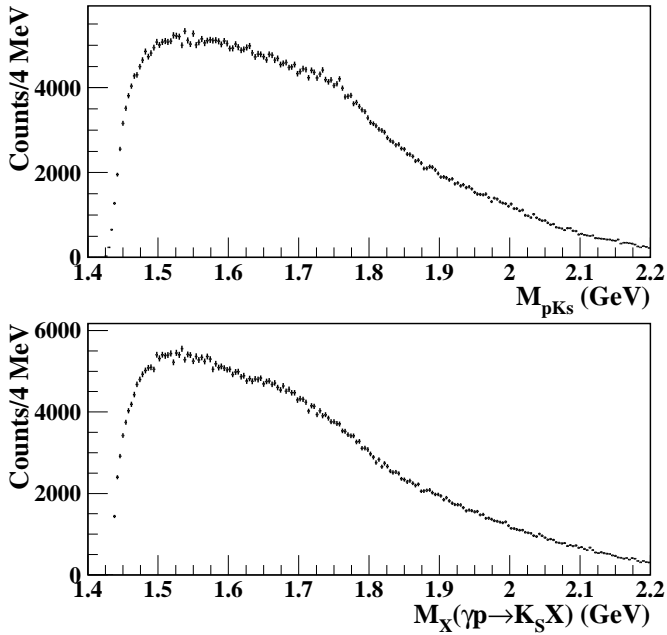}
\caption[]{The $pK_S$ invariant mass (top) and $K_S$ missing mass (bottom) for the reaction $\gamma p \to \bar K^0 K^0p$ 
 after all cuts,  obtained by integrating over all measured
photon energy and angles. Shoulders corresponding to excited $\Sigma$ states are visible in both spectra.}
\label{fig:nkp_spectrum2}
\end{figure}

\begin{figure}
\vspace{8.0cm} 
\includegraphics{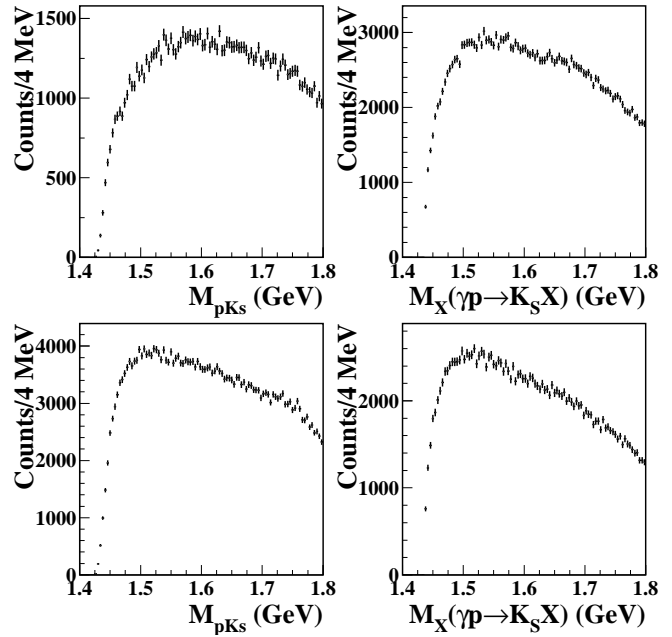}
\caption[]{The $pK_S$ invariant mass (left) and  $K_S$ missing mass spectra (right) for the reaction $\gamma p \to \bar K^0 K^0p$
  for different angular ranges:
$\cos\theta^{CM}_{\bar K^0}<0$ (top), and  $\cos\theta^{CM}_{\bar K^0}>0$ (bottom).}
\label{fig:nkp_angcut2}
\end{figure}

The selected final state contains two neutral kaons, one detected and one missing,  therefore a possible $\Theta^+$ peak 
can show up in  two ways: in the  invariant mass spectrum of the $pK_S$ system  or in the missing mass spectrum  
of the  detected $K_S$.
Fig.~\ref{fig:nkp_spectrum2} shows the two spectra after all cuts:  
both of them are smooth and structureless. In particular no evidence for 
a peak or an  enhancement is observed at masses close to 1540 MeV, where signals 
associated with the $\Theta^+$ were previously reported. 
To enhance a possible resonance signal not visible in the integrated spectra, 
we assumed  the  two body reaction $\gamma p \to \bar K^0 \Theta^+(1540)$
and selected different $\bar K^0$ center-of-mass angle intervals. The  $\bar K^0$
angle was calculated using the reconstructed kinematic variables of the missing kaon  in the first case
and the measured kinematic variables  of the observed kaon in the second case.
Fig.~\ref{fig:nkp_angcut2} shows the $pK_S$ invariant mass  and the 
$K_S$ missing mass spectra  for forward and backward
$\cos\theta^{CM}_{\bar K^0}$ ranges separately.
When the  $pK_S$ system is considered, Monte Carlo studies  showed that the 
CLAS acceptance is maximum at  forward $\theta^{CM}_{\bar K^0}$ and therefore
complementary to what we found for the $\Theta^+ \to n K^+$ decay mode (as seen in Sec.~\ref{ssec:reac1}).
No structures were found in any of the spectra when specific angular ranges were selected.
\begin{figure}
\vspace{7.5cm} 
\includegraphics{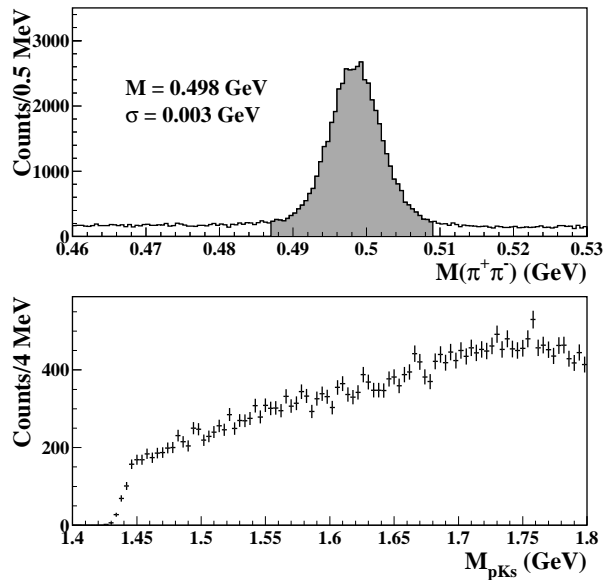}
\caption[]{Effect of the  $K_S$ decay length cut  on the analysis.
Top: $\pi^+ \pi^-$ invariant mass and the $K_S$ peak (as in Fig.\ref{fig:pid2}-top). 
Bottom: $pK_S$ invariant mass (as in Fig.\ref{fig:nkp_spectrum2}-top).}
\label{fig:k0_vcut}
\end{figure}

As shown in Fig.~\ref{fig:pid2} the $K_S$ peak sits over a large background mainly related 
to multi-pion production. A cleaner sample is obtained
by applying a cut on the $K_S$ decay length:
in fact, due to the sizeable  $K_S$ mean life (c$\tau\sim$ 2.68 cm),
its decay vertex ($K_S \to \pi^+ \pi^-$) is detached from the primary production vertex 
($p \bar K^0 K^0$). Taking into account the vertex resolution of the CLAS detector 
($\sim$ 0.3 cm) and the $K_S$ c$\tau$, we applied a  3 cm  decay length cut, obtaining 
the  mass spectra of Fig.~\ref{fig:k0_vcut}.
Despite the use  of a cleaner $K_S$ sample (top panel),
no structures are present in the $pK_S$ invariant mass spectrum (bottom panel), 
confirming the results reported  above.
The $K_S$ decay length cut improves the signal-to-background ratio for the $K_S$ identification, cleaning the data sample from 
the multipion contamination. On the other side it reduces the $K_S$ yields by  
almost a factor five and, due to the strong correlation of the  $K_S$ mean life with the kaon momentum, 
it results in a momentum-dependent cut on the  $K_S$ sample, difficult to reproduce by  Monte Carlo simulation.
This also distorts the $pK_S$ invariant mass spectrum at low values where a possible resonance 
is more likely produced. For these reasons, the upper limit for this decay mode are evaluated 
without the vertex cut.
\begin{figure}
\vspace{15.cm} 
\includegraphics{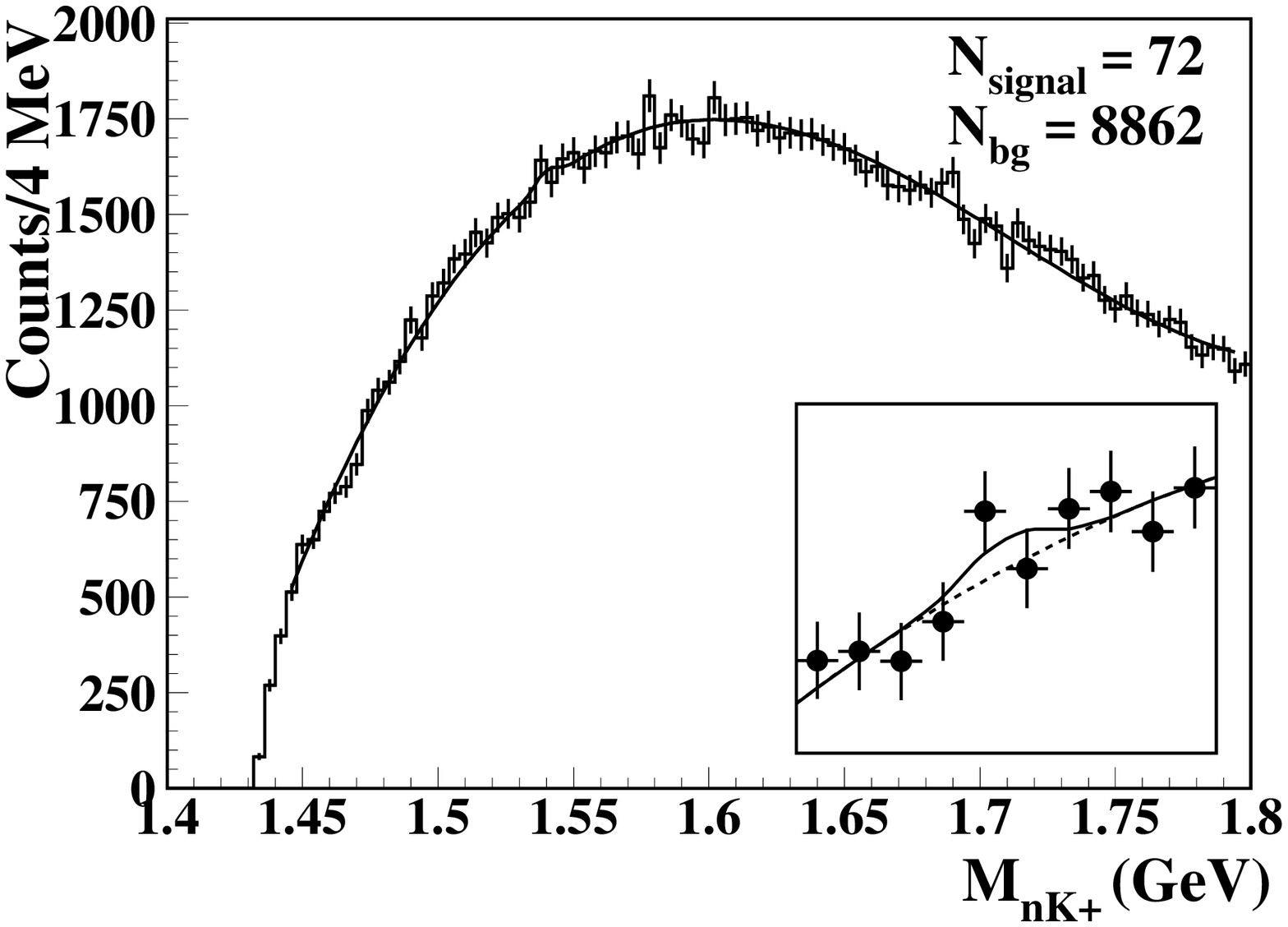}
\includegraphics{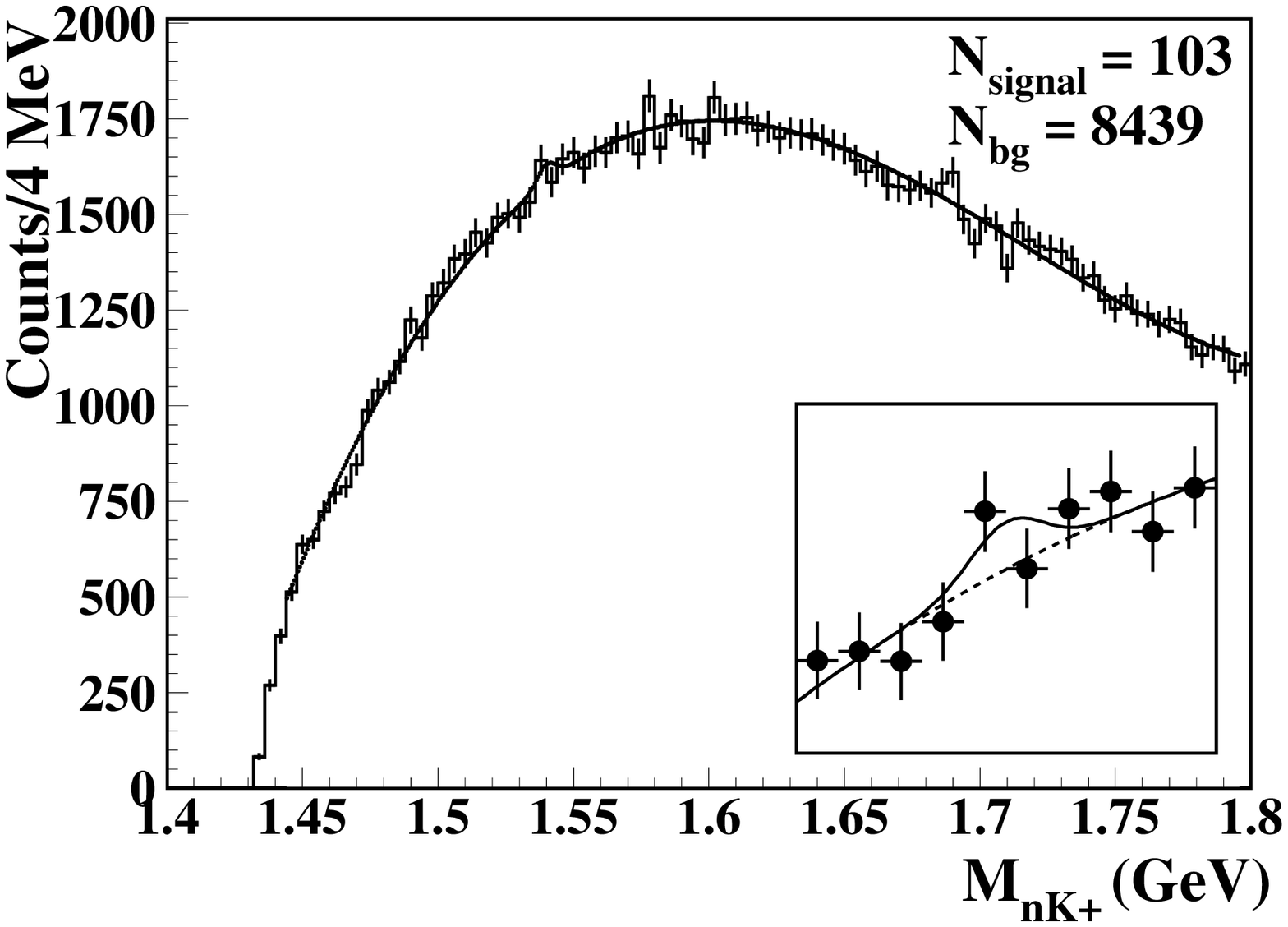}
\includegraphics{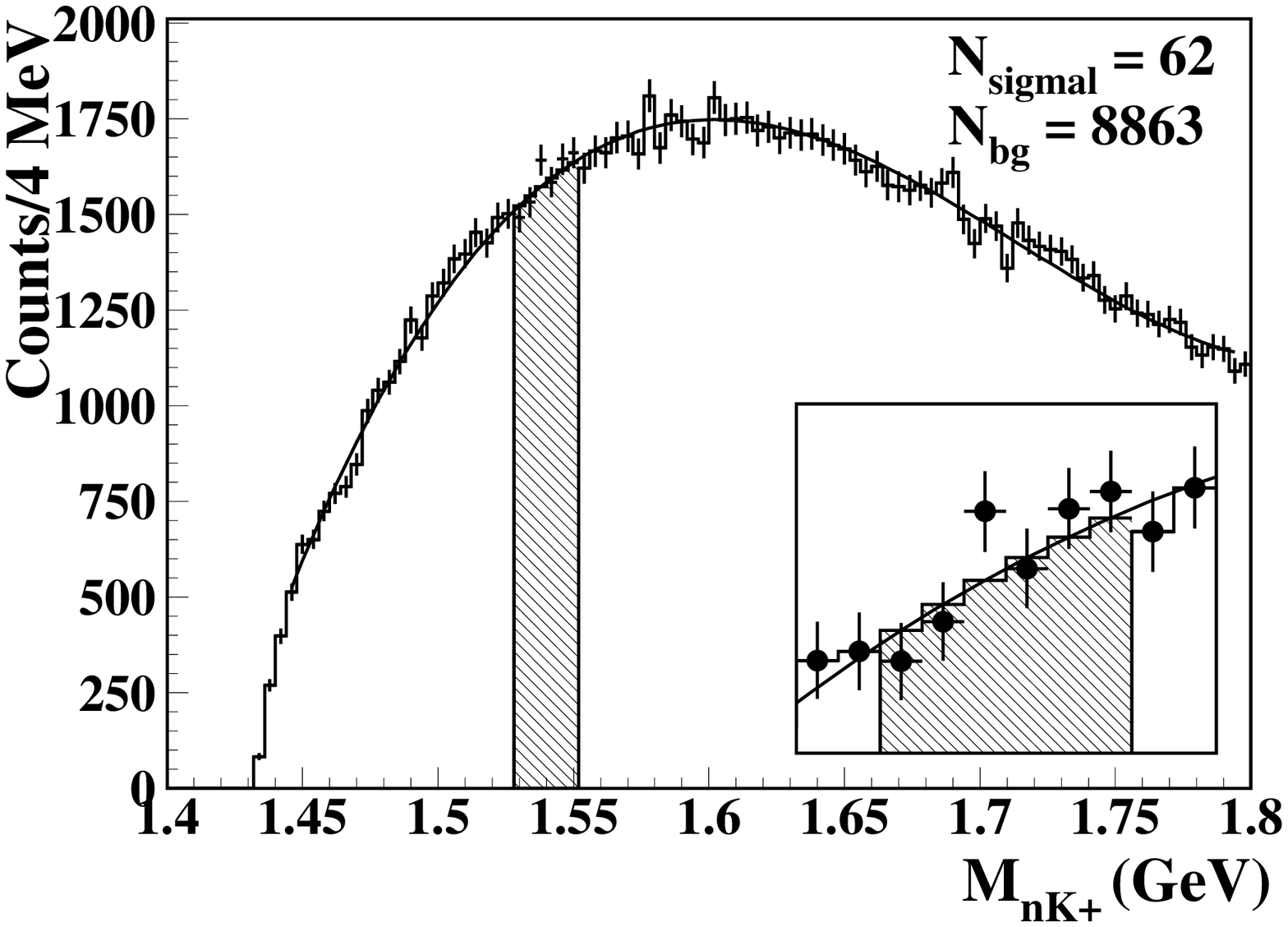}
\caption[]{The $nK^+$ invariant mass spectrum for the reaction  $\gamma p \to  \bar K^0 K^+ n$ 
fitted for a $\Theta^+$ mass of 1540 MeV using the three different methods described in the text: 
from top to bottom the $\chi^2$ fit of the histograms, the likelihood fit of the unbinned events, 
and the background fit with the exclusion of the $\Theta^+$ mass window (indicated as hatched area). The obtained yields for signal and background are  shown on top of each plot. Insets show the zoom of the mass region 1520--1560 MeV.}
\label{fig:fits}
\end{figure}

\subsection{Upper limit on the $\Theta^+$ yields }\label{sec:fit}
Since no signal was observed,
an upper limit on the $\Theta^+$ yield was extracted and 
transformed to an upper limit on the production cross section 
in each reaction channel.
In this section we discuss in detail the procedure adopted  for the 
channel $\gamma p \to \bar K^0 K^+ n$ and, since the procedure is the same, we  summarize the results
for the channel   $\gamma p \to \bar K^0 K^0 p$.

The $\Theta^+$ was assumed to be a narrow peak over a smooth background.
The $nK^+$ invariant mass spectrum was fit to 
the sum of a  Gaussian-shape resonance  and a fifth-order polynomial representing the background.
The resonance  position was varied from 1520 to 1600 MeV in 5 MeV steps while the width 
was kept fixed assuming the dominance of the experimental resolution over the intrinsic 
width as suggested from recent analyses of $KN$ scattering data \cite{kn}.

\begin{figure}
\vspace{10.0cm} 
\includegraphics{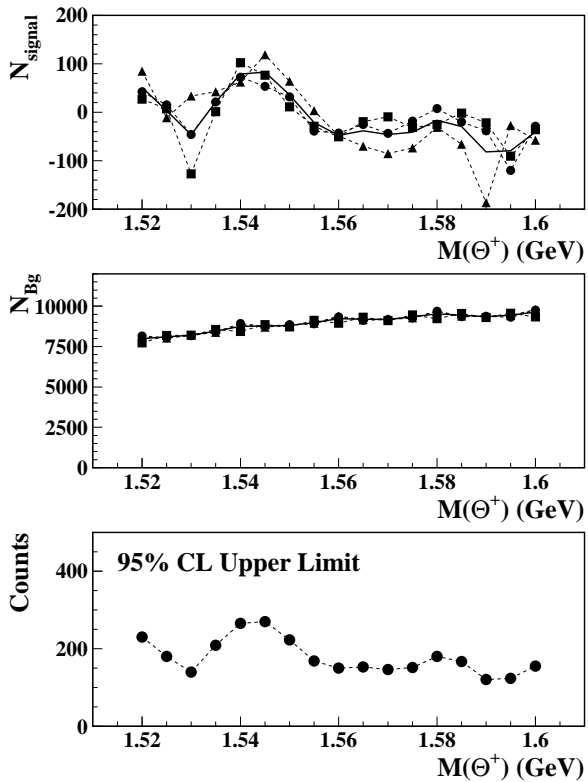}
\caption{Signal (top) and background (middle) yields obtained with the three fit procedures 
for the reaction  $\gamma p \to \bar K^0  K^+ n$:
$\chi^2$ fit of the histograms (circles),  likelihood fit of the unbinned events (squares),
and  background fit with the exclusion of the $\Theta^+$ mass window (triangles). The dashed lines are to guide the eye.
The three results were combined together taking the average of the event yields (solid line) 
and transformed in 95\% C.L. upper limit  shown in the  bottom panel.}
\label{fig:up_mass}
\end{figure}

The $\Theta^+$  width, $\sigma_{\Theta^+}$, was derived by means of a Monte Carlo simulation.  A zero-width resonance was generated and 
projected  over the CLAS detector, applying the same analysis chain used to process the data. 
A width of $\sigma_{\Theta^+}$ of 3--4 MeV was obtained, which weakly depends on the photon energy and the $\bar K^0$ emission angle.
To check the consistency of the experimental resolution obtained 
from the Monte Carlo simulations, the same procedure was applied to the reactions 
$\gamma p  \rightarrow  K^+ \Sigma^+ \pi^-$, and $\gamma p  \rightarrow  K^+ \Sigma^- \pi^+$ where 
a direct comparison to the data is possible (see Sec.~\ref{ssec:reac1}). The width of the two hyperons derived from the simulations 
was found to be compatible  with the measured values. 
For each  value of the $\Theta^+$ mass, the resonance and background yields 
were extracted using three different fit procedures:
(1) a $\chi^2$ fit of the mass distribution binned in 4 MeV channels;
(2) a likelihood fit of the unbinned $nK^+$ spectrum;
(3) as in case (1) but with  the background function  being fit after excluding the signal region 
defined as $\pm 3 \sigma_{\Theta^+}$  around the selected mass value.
In all cases the background yield was obtained by
integrating the polynomial function over 
$\pm 3 \sigma_{\Theta^+}$ of the selected $\Theta^+$ mass value. In  methods (1)  and (2) above, 
the signal yield was obtained as the integral of the resulting Gaussian, 
while in  method (3)  it was obtained as the 
difference between the number of observed events and the background integrated 
over $\pm 3 \sigma_{\Theta^+}$ of the chosen $\Theta^+$ mass value.
The same procedure was then repeated subdividing the data into 16 $\cos\theta^{CM}_{\bar K^0}$ 
bins producing binned spectra.
The results of the three methods applied to the integrated spectrum are shown in Fig.~\ref{fig:fits}. 
In general, the signal yields obtained with the three procedures are compatible with zero within 
1 or 2 sigma, confirming that no evidence for $\Theta^+$ production
is observed in the mass range 1.52--1.6 GeV. The results of the binned  $\chi^2$ fits and un-binned likelihood fit are in good agreement with
each other showing that the binning effects are small. The measured yields and the background are shown in 
the top and middle panels of Fig.~\ref{fig:up_mass}.
In general the results of the three methods ($\chi^2$, likelihood, and $\chi^2$ without the $\Theta^+$ mass window) 
are consistent with each other  as expected by the dominance of the background over the signal yield. They 
were  combined by  taking the average of the event yields, for both signal and background,
in the conservative assumption of totally correlated measurements.
The averaged yields were transformed into upper limits of the $true$ $\Theta^+$ yield using 
the Feldman and Cousins approach~\cite{fc99}.
This method determines proper confidence level boundaries for small signals over a  background taking into account 
external constraints (e.g. the $true$ yield is constrained to be positive).
In addition it decouples the goodness-of-fit confidence level from the confidence level interval.
The resulting upper limit at 95\% C.L. is shown in the bottom panel of Fig.~\ref{fig:up_mass}; it is almost flat with the
maximum value  around 1545 MeV.

On average, the upper limit at 95\% C.L. on the yields is $N_{\Theta^+} \sim 220$.
The ratio of the yield of ${\Theta^+}$ to the ${\Lambda^*(1520)}$ has also been obtained:
$\frac{N_{\Theta^+}}{N_{\Lambda^*(1520)}}\sim 220/99000 \sim 0.22\%$ 
(the two yields are not corrected for the CLAS efficiency).  
\begin{figure}
\vspace{10.0cm} 
\includegraphics{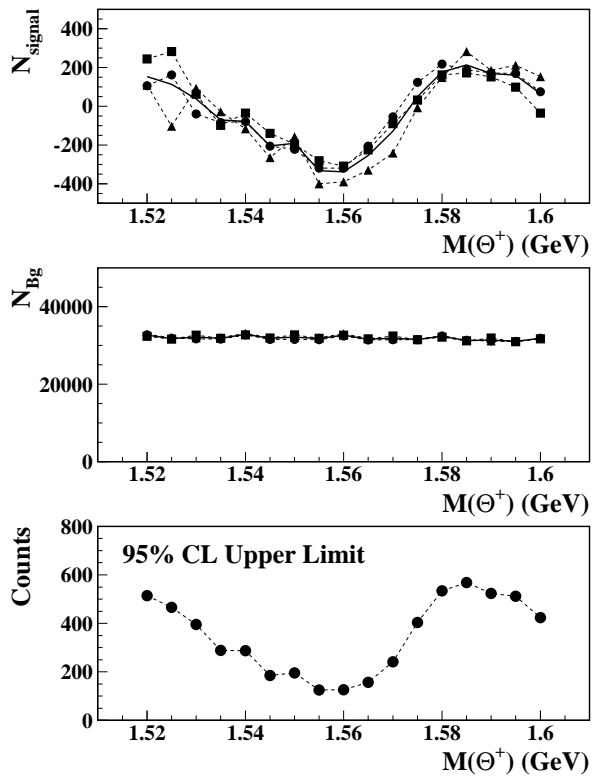}
\caption{The same as in Fig.~\ref{fig:up_mass} for the  $\gamma p \to \bar K^0  K^0  p$ channel.}
\label{fig:inc_yields}
\end{figure}
\begin{figure}
\vspace{7.1cm} 
\includegraphics{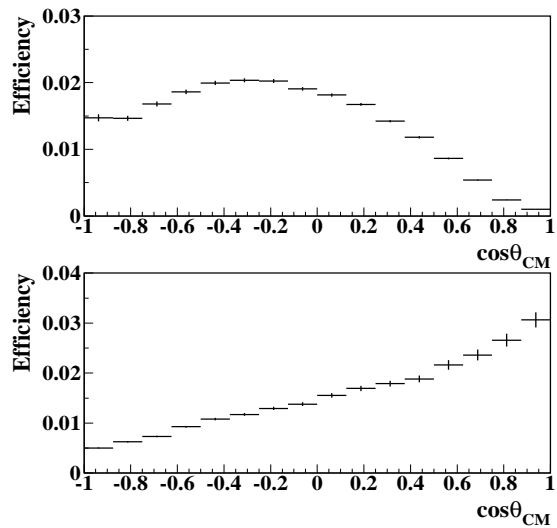}
\caption{Binned CLAS efficiency for the reaction $\gamma p \to \bar K^0 \Theta^+$.
Top: $\Theta^+ \to n K^+$ assuming the $t-$exchange dominance hypothesis;
bottom: $\Theta^+ \to p K^0$ assuming the calculation of 
Ref.~\cite{oh04} with no $K^*$-exchange. 
These two correspond to the `worse case  scenario' for each decay mode respectively.}
\label{fig:eff_diff}
\end{figure}
The same procedure  was repeated for the  $\gamma p \to \bar K^0  K^0  p$  channel
to derive  an upper limit at 95\% C.L. on the yield
as a function of the $p K_S$ invariant mass and on the differential cross section assuming a $\Theta^+$ mass of 1540 MeV.
The upper limit was derived from the $pK_S$ mass spectrum only 
(the $K_S$ missing mass spectrum was ignored because it is correlated to the $pK_S$ mass spectrum being built from the same event sample).
The better CLAS acceptance for the proton and the $K_S$ coming  from the $\Theta^+$ decay
leads to  a complete angular coverage complementary to the $\gamma p \to \bar K^0  K^+ n$ channel.
The $\Theta^+$ was searched for as a narrow peak by fitting  the $p K_S$ mass spectrum with a Gaussian 
curve with $\sigma_{\Theta^+}\sim$4  MeV inferred from dedicated Monte Carlo simulations,
plus a fifth-order polynomial representing a smooth background. The mass region 1520--1600 MeV was scanned
in 5 MeV steps.
To derive the yields of a possible resonance and associated  background the three fit procedures described 
for the $\gamma p \to \bar K^0  K^+ n$  channel
were applied to the integrated spectrum and  the $\cos\theta^{CM}_{\bar K^0}$-binned spectra for a fixed $\Theta$ mass of 1540 MeV.
The three results were combined  taking the average of the event yields. These were  transformed into 
95\% C.L. upper limits. 
Figure~\ref{fig:inc_yields} shows the comparison of the 3 fitted yields as a function of the $pK_S$ invariant mass and the resulting 95\% C.L. upper limit.
\begin{figure}
\vspace{7.5cm} 
\includegraphics{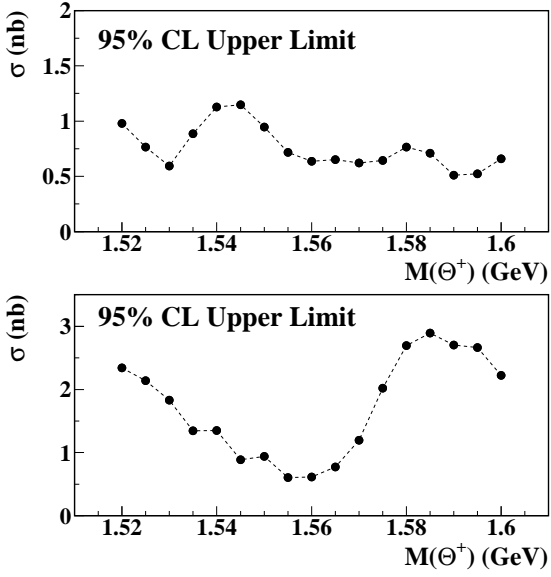}
\caption{The 95\% Confidence-Level upper limit on the total cross section for the reaction
$\gamma p \to \bar K^0 \Theta^+$  with
$\Theta^+ \to n K^+$ (top) and $\Theta^+ \to p K^0$ (bottom). The dashed line is to guide the eye.} 
\label{fig:xsec1}
\end{figure}

\subsection{Upper limits on the $\Theta^+$ Cross Section}\label{ssec:ul}
The 95\% C.L. upper limits on the yield  described in the previous section were then transformed into limits 
on the $\Theta^+$ production cross section according to the following formula:

\begin{equation}\label{eq:xsec1}
\sigma_{nK^+} = \frac{N_{nK^+}}{\mathcal{L}\;\;\epsilon_{nK^+}\;\;b_{nK^+}}
\end{equation}
\begin{equation}\label{eq:xsec2}
\sigma_{pK^0} = \frac{N_{pK^0}}{\mathcal{L}\;\;\epsilon_{pK^0}\;\;b_{pK^0}}
\end{equation}

\noindent
where $N$ is the 95\% C.L. limit on the $\Theta^+$ yield, $\epsilon$ is the CLAS efficiency, $\mathcal{L}$ is the 
integrated luminosity, $b$ is  the branching ratio for the $\Theta^+$ decay, and subscripts indicate the decay mode of  
the $\Theta^+$. Each branching ratio is assumed  to be 50\%.
The branching ratios for  neutral kaon decay to $K^0 \to K_S \to \pi^+\pi^-$ ($50\% \cdot 68\%$) were included in the CLAS efficiency.

The luminosity was obtained as the product of the target density 
and length, and the incoming photon flux measured during the experiment 
corrected for data-acquisition dead time. When the  $\Theta^+$ mass was varied from 1.5 to 1.6 GeV, the production threshold in beam energy
moved from  1.65 to 1.85 GeV. The photon flux used in the cross section estimate was calculated accordingly.
\begin{figure}
\vspace{7.5cm} 
\includegraphics{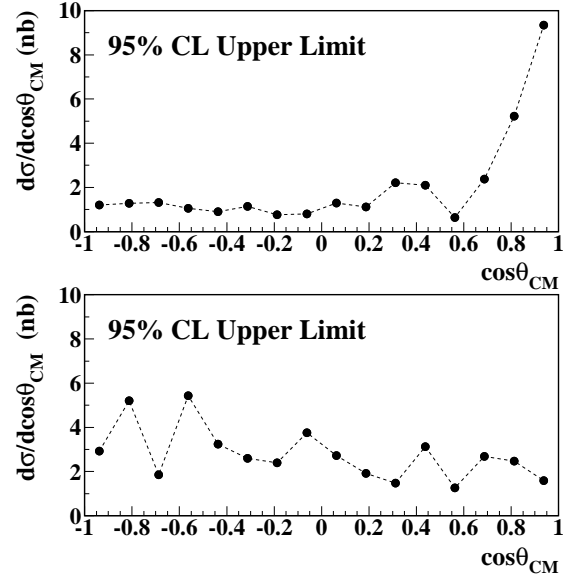}
\caption{The 95\% Confidence-Level upper limits on the  differential cross section $d\sigma/d\cos\theta^{CM}_{\bar K^0}$
for the reaction $\gamma p \to \bar K^0 \Theta^+(1540)$ obtained from the two decay modes:
$\Theta^+(1540) \to n K^+$ (top) and $\Theta^+(1540) \to p K^0$ (bottom). The dashed line is to guide the eye.} 
\label{fig:xsec2}
\end{figure}

\begin{figure}[h]
\vspace{6.5cm} 
\includegraphics{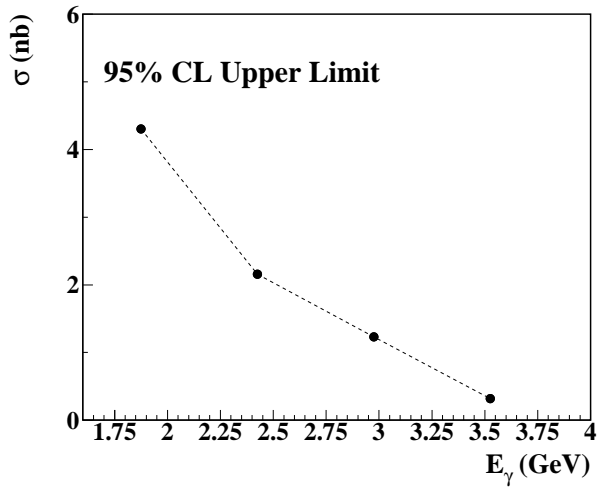}
\caption{The 95\% Confidence-Level upper limits on the total cross section 
as a function of $E_\gamma$ for the reaction $\gamma p \to \bar K^0 \Theta^+(1540)$ with  $\Theta^+ \to\ n K^+$. The dashed line is to guide the eye.} 
\label{fig:xsec3}
\end{figure}

\begin{figure}
\vspace{9.0cm} 
\includegraphics{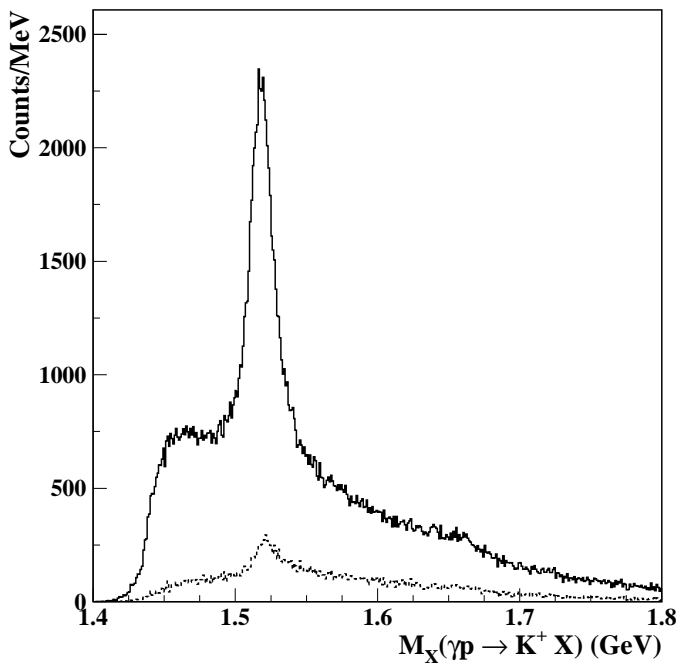}
\caption[]{The CLAS  $K^+$ missing mass showing  the $\Lambda^*(1520)$ peak before (solid) and after (dashed) cutting on 
$\cos \theta^{CM}_{\bar K^0}>$0.5.}
\label{fig:fig110}
\end{figure}

The CLAS detection efficiency was obtained by means of detailed Monte Carlo studies.
The reaction $\gamma p \to \bar K^0 \Theta^+$ and subsequent $\Theta^+$ decay to  $n K^+$
and  $p K^0$ was generated 
assuming  different  production mechanisms: $t-$exchange
dominance (the $\bar K^0$ mainly  produced at forward angles in the 
center-of-mass system), $u-$exchange dominance ($\bar K^0$  at backward angles), 
$\cos\theta^{CM}_{\bar K^0}$ uniformly distributed, and 
using the predictions of the model in Ref. \cite{oh04}. 
For the $t-$exchange hypothesis we assumed
the same angular distribution 
as for the reaction $\gamma p \to \Lambda^*(1520) K^+$, which exhibits the typical 
$t-$channel forward peaking behavior  (approximately an exponential with a slope of $-2.5$ GeV$^{-2}$).
The $u-$exchange distribution was generated the same way except that  the center-of-mass
angles of the $\bar K^0$ and $\Theta^+$ were interchanged.

For $\gamma p \to \bar K^0 \Theta^+$, $\Theta^+ \to K^+ n$,
the CLAS overall detection efficiencies obtained 
with  the various production mechanisms vary between 1\% for the $t-$exchange  hypothesis 
to 1.8\% for the angular distribution of Ref.~\cite{oh04} when no $K^*$ exchange process 
is included. 
As a function of $\theta^{CM}_{\bar K^0}$ all the different hypotheses gave a comparable efficiency: almost flat 
from 180$^\circ$ to 90$^\circ$ (about 2$\%$) and then smoothly dropping at forward angles. 
For $\gamma p \to \bar K^0 \Theta^+$,  $\Theta^+ \to  K^0 p$,  the efficiency varied between 1\% for the angular distribution 
of Ref.~\cite{oh04} with no $K^*$ exchange process to  1.8\% for the  $t-$channel hypothesis, with an angular dependence complementary 
to the other channel (smoothly increasing from backward to forward angles). 
For each branch, the model that yielded the lowest efficiency was chosen for conservatism. The resulting
efficiencies are shown in Fig.~\ref{fig:eff_diff}.

The upper limits on the total cross
sections as a function of the $\Theta^+$ mass were obtained independently for each decay mode
using the model assumptions described above. The results are shown in Fig.~\ref{fig:xsec1}.
For $\Theta^+ \to n K^+$ a  95\% C.L. upper limit of 1.0 nb was found for $M_{\Theta^+}$ = 1540 MeV.
The corresponding limit for $\Theta^+ \to p K^0$ was 1.3 nb.

The 95\% C.L. upper limit on the $\Theta^+$(1540) differential cross section
 $d\sigma/d\cos\theta^{CM}_{\bar K^0}$  is shown  Fig.~\ref{fig:xsec2}, using
the same assumption on the production mechanisms as for the evaluation of the  upper limit on the total cross section.
However, for this quantity 
no significant difference was found when  the other hypotheses were
 used in the efficiency evaluation.
For the $n K^+$ decay mode the cross section limit remains within 1--2 nb for most of the angular range and rises at forward angle due
to the reduced CLAS acceptance. For the $p K^0$ decay mode, the cross section limit is 
 within 2--5 nb over the entire angular range.

Finally, for the reaction $\gamma p \to \bar K^0 \Theta^+(1540)$ with   $\Theta^+ \to n K^+$,
the $\Theta^+$(1540) total cross section
upper limit as a function of the photon energy is shown in  Fig.~\ref{fig:xsec3}; 
the behavior reflects the CLAS efficiency, which  is reduced at low energy near the $\Theta^+$(1540) production threshold,
 and then  increases with energy,  resulting in a better limit for higher energies.
\begin{figure}
\vspace{9.0cm} 
\includegraphics{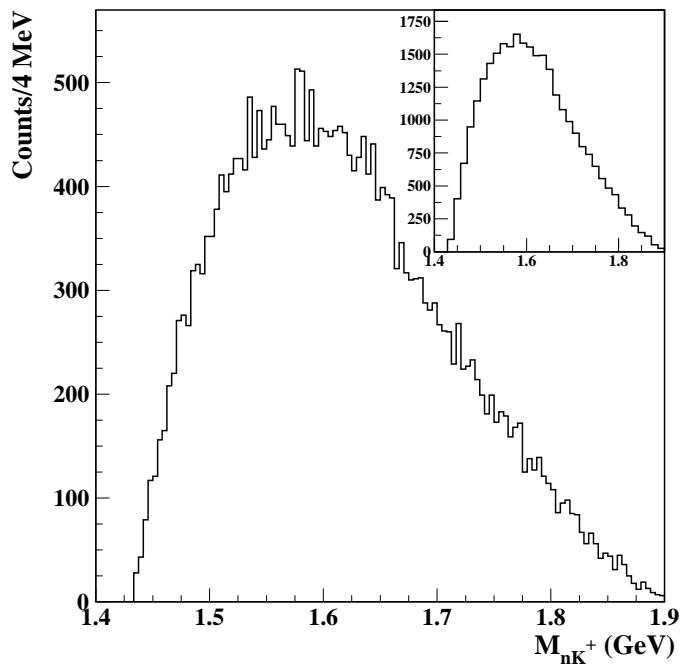}
\caption[]{The CLAS $nK^+$ invariant mass with all the cuts as in Ref.~\cite{saphir}. 
The inset show the same spectrum with the binning used by the authors of the same reference.}
\label{fig:fig111}
\end{figure}

\section{\label{sec:comp}Comparison with existing data}
Our result for the reaction
$\gamma p \to \bar K^0 \Theta^+ \to \bar  K^0 K^+ n$
is in clear disagreement with the findings of Ref.~\cite{saphir}
which reported a $\Theta^+$ signal  of 55 events at a mass of 1540 MeV corresponding to an 
estimated total cross section of 50 nb.

In order to better compare our data with this experiment, the  
kinematic cuts described in Ref.~\cite{saphir}  were  applied to the present data. 
The  photon energy was limited to be below 2.6 GeV and  a possible 
forward peaked production of the $\Theta^+$ was selected by applying the angular cut 
$\cos\theta^{CM}_{\bar K^0}>0.5$. This cut also reduces the  hyperon production yield.
The effect is shown in 
Fig.~\ref{fig:fig110} where the  missing mass of the $K^+$ is shown
before and after the angular cut: the $\Lambda^*(1520)$ is clearly suppressed.
As a result, no hyperon rejection cuts were applied.

The  $\bar K^0$  missing mass spectrum, after all cuts, is shown in Fig.~\ref{fig:fig111}
with two different bin sizes reflecting the CLAS and the SAPHIR resolutions.
There is no evidence of a $\Theta^+$ peak.

Applying the same procedure described above, we evaluated a 95\% confidence level limit 
on the $\Theta^+(1540)$ yield with SAPHIR selection cuts of $90$ events. 

To derive the  $\Lambda(1520)$ yield, the  $K^+$ missing mass spectrum obtained before the angular cut was fit 
by  a Breit-Wigner function plus a second-order polynomial background, 
with the same procedure described in Sec.~\ref{ssec:reac1}, obtaining (57,000$\pm$5,500) $\Lambda^*(1520)$s. 
This number  is to be  compared to 630$\pm$90 reported by SAPHIR Ref.~\cite{saphir}.
The ratio between  observed $\Theta^+$ and $\Lambda(1520)$ in this experiment is $\sim 0.16\%$ differing
by more than a factor 50 from the value quoted in Ref.~\cite{saphir}. All the yields reported above are notcorrected
for detector acceptances.

No results were published  on the search of the  $\Theta^+$ in the exclusive reaction  $\gamma p  \to \bar K^0 K^0 p$
and therefore no comparison is possible for this channel.

\section{\label{sec:syst}Systematic errors}
In the evaluation of the upper limit we have considered the following sources of systematic errors:
determination of the mass resolution for the $\Theta^+$ resonance,
determination of the signal and background yields from the mass spectra,
evaluation  of the CLAS efficiency, 
detector inefficiencies, photon flux  normalization, and
dependence on the analysis procedure. 
The first three sources were already included in the quoted upper limit as explained in the following subsections. 
In particular, the model dependence in evaluating the CLAS efficiency was estimated by comparing the results obtained using 
different models for the production cross section. 
The resulting efficiencies differ by a maximum of a factor two. The upper 
limits were estimated using always the worse case scenario.
The remaining  sources of systematic uncertainty, summarized in Table~\ref{table:sys}, 
result in an overall systematic error of $\epsilon =  25\%$  accounted for by
multiplying the upper limit by (1+$\epsilon$). 

\begin{table}[!h]
\caption{Systematic errors on the upper limit evaluation.}
\label{table:sys}
\begin{tabular}{| c | c | c |}
\hline
Source & Error (\%)\\
\hline\hline
Detector inefficiencies  & 10  \\
Photon flux normalization & 10  \\
Analysis procedure  & $<$20  \\
\hline
\end{tabular}
\end{table}

In the following, the different contributions to the total systematic error are discussed in more detail.

\subsection{Mass resolution and evaluation of signal and background yields}
As discussed in Sec.~\ref{sec:fit} the mass resolution for a narrow resonance 
was estimated from Monte Carlo simulation; the reliability of simulations in reproducing the CLAS resolution 
was tested comparing the observed resolution for known narrow resonances and a maximum discrepancy of 
20\% was found. The resolution for the $\Theta^+$ peak extracted from Monte Carlo was therefore rescaled by a 
factor 1.2 to account for this.

The comparison of different fitting procedures provides an estimate of the associated systematics. 
The upper limits were derived combining the results of the three fits, including therefore an estimate of the associated error.
\begin{figure}[h]
\vspace{7.5cm} 
\includegraphics{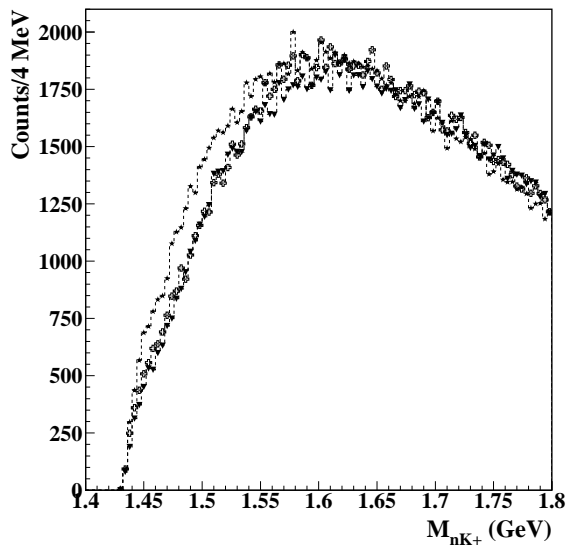}
\caption[]{The final $nK^+$ spectra  for the reaction  $\gamma p \to  \bar K^0 K^+ n $ obtained by the three independent  analyses.} 
\label{fig:spectra}
\end{figure}

\subsection{Detector inefficiencies  and normalization}
As a check of the accuracy of the CLAS detector simulations and photon flux normalization, 
the differential and the total cross section for several known 
reactions were derived from  this data set. 
Due to the high multiplicity of charged particles in the final state in our data (similar to  $\bar K^0 K^+ n$ and $\bar K^0 K^0 p$)
and the existence of precise measurements that can be taken as a reference, 
 the reactions $\gamma p \rightarrow p \omega$ and  
$\gamma p \rightarrow K^+ \Lambda(1116)$ 
were used as a test of the different ingredients used in our analysis. Moreover,
the measurement of different final states of the same reaction, such as $p \pi^+ (\pi^- \pi^0)$ and 
$p \pi^+ \pi^- (\pi^0)$ for the $\omega$ photoproduction and  $K^+ p (\pi^-)$ and  $K^+ p \pi^-$
for the $K^+ \Lambda(1116)$ channel, were also used to test
the hardware trigger, the photon flux normalization and the procedure to extract the CLAS efficiency. 
The differential and the total cross sections extracted from this data set agree with each other and with
the world data within the experimental error,  verifying that  
the different steps in the analysis are in control at the 15$\%$ level. 

To directly check our ability to observe the final state involved 
in the pentaquark decaying into the $nK^+$ final state,
the cross section for the  reaction $\gamma p  \rightarrow  K^+ \Lambda^*(1520) \rightarrow  K^+ \bar K^0 n$ was also extracted. As shown above,
the $\Lambda^*(1520)$ peak is clearly visible, and the sample of 100,000 $\Lambda^*$'s made possible 
an analysis  deriving both the differential and the total cross sections.
The $\Lambda^*(1520)$ cross sections we obtained were compared with data from the
SAPHIR~\cite{lstar-saphir} and NINA~\cite{lstar-nina} collaborations. 
We are in good agreement with the SAPHIR results but find a much lower cross section than that reported by NINA.
The results of all these measurements will be reported elsewhere.
\begin{figure}
\vspace{7.5cm} 
\includegraphics{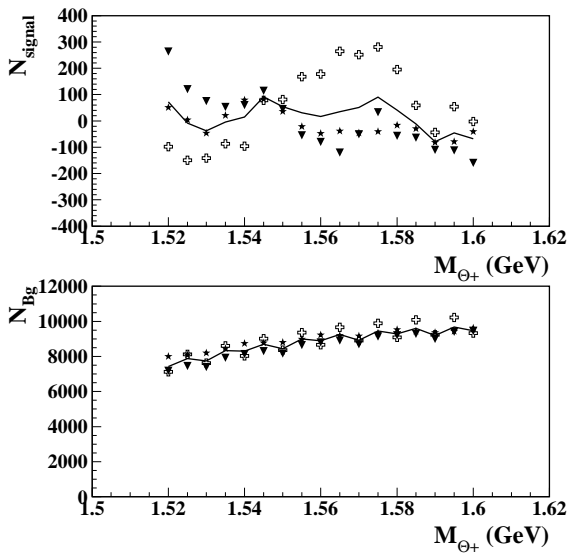}
\caption[]{
Signal (top) and background (bottom) yields 
obtained by the three independent analyses for the reaction $\gamma p \to \bar K^0 K^+ n$.
Different symbols refer to the different analyses.
The solid line shows the average of the three results.}  
\label{fig:yield}
\end{figure}

\subsection{Dependence on the analysis procedure }\label{sec:ind}

Three independent analyses were conducted both on the reactions of interest for the pentaquark
search and on reference reactions as $\gamma p \to \Lambda^*(1520) K^+ \to  \bar K^0 K^+ n$.
This enabled  an evaluation of the systematic errors associated with the analysis procedures 
and provided a cross check on the results.
The three analyses were not totally independent since they all used the same raw data and the same basic corrections
to the measured kinematic quantities such as the energy loss.
However, they used different particle identification
schemes, different detector calibration procedures (for both the tagger system and the CLAS spectrometer),
different Monte Carlo simulations to evaluate the CLAS efficiency, and different fit procedures to extract yields.

The  $\Lambda^*(1520)$ differential and total cross
sections obtained by the three analyses  were found to be
consistent with each other at the  10$\%$ level and agree with the SAPHIR measurement at the same level.

\begin{figure}
\vspace{7.5cm} 
\includegraphics{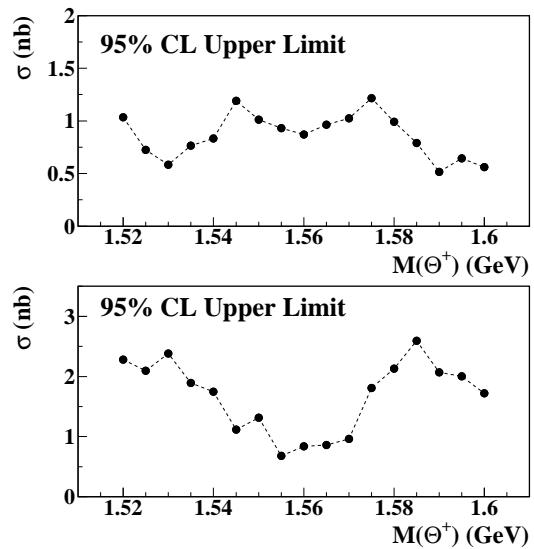}
\caption[]{The  95\% C.L. upper limit on the cross section for the reactions 
$\gamma p \to \bar K^0 \Theta^+ \to  \bar K^0 K^+ n$ (top) and 
$\gamma p \to \bar K^0 \Theta^+ \to  \bar K^0 K^0 p$ (bottom)
derived by the combination of the three analyses. The dashed line is to guide the eye.
As explained in the text, the three analyses were largely independent.} 
\label{fig:yield-final}
\end{figure}

For the reaction $\gamma p \to   \bar K^0  \Theta^+ $,
all  three analyses agreed on the main conclusion:
the  $nK^+$ and $pK_S$ spectra are smooth and structureless and, in particular, no signal is observed
at 1540 MeV where the  $\Theta^+$ pentaquark has been widely reported.

The  $nK^+$ mass spectra for the reaction $\gamma p \to \bar K^0K^+n$   obtained by the three analyses
are shown in Fig.~\ref{fig:spectra}.
The difference in the shape of the spectra  is mainly related to the different particle identification
schemes adopted by the three groups while the small differences in the bin-to-bin fluctuations are due to the
different kinematic corrections applied in the  analyses.
Signal and background yields  as a function of the $\Theta^+$ mass  resulting
from the fit of the three spectra are shown in Fig.~\ref{fig:yield}.
The three analyses are consistent at the 10\% level on the
background estimate while they differ in the event yield evaluation.
This discrepancies, reflecting the different choices in the analysis procedures,
provide an estimate of the systematic error associated to the extraction of the upper limits.

The same comparison was performed for the $\gamma p \to \bar K^0 K^0 p$ channel with similar results.

\section{\label{sec:ro} Results}

\subsection{Upper limits on the  $\Theta^+$ production cross section}
The independent analyses  were combined, taking the average of the event yields, and
transformed into the 95\% C.L. upper limit on the yields with the Feldman and Cousins procedure.
They were  then 
transformed  into the  95\% C.L. on  the cross section using the CLAS efficiency evaluated 
in the most conservative scenario. Results are shown in Fig.~\ref{fig:yield-final}.

For the reaction $\gamma p \to \bar K^0 \Theta^+$ with $\Theta^+ \to n K^+$
the upper limit at 95\% C.L. on the $\Theta^+$ production cross section varies between 0.5 nb and 1.3 nb as a 
function of the $nK^+$ invariant mass with a value of $\sim$0.8 nb at 1540 MeV.
For the reaction $\gamma p \to \bar K^0 \Theta^+$ with $\Theta^+ \to p K^0$
the upper limit at 95\% C.L. on the $\Theta^+$ production cross section varies between 0.5 nb and 2.5 nb as a 
function of the $pK_S$ invariant mass with a value of $\sim$1.5 nb at 1540 MeV.
The  results for the two decay modes are similar in value  and  set
stringent upper limits on the models which predict these long-lived pentaquark states.

So far the two reaction channels were  independently analyzed. Assuming  they result from the 
two possible decay modes of the $\Theta^+$, they can be combined  to give a single  upper limit.

As shown in the previous sections, we estimated a signal yield $S_i$ (with $i=$ 1 and 2, corresponding to
$nK^+$ and $pK^0$ branching mode), e.g. the area of a Gaussian of fixed width and
fixed mass fit to the mass histogram, and a background yield $B_i$, e.g. the  polynomial background.

These were corrected for the detection efficiency and luminosity to obtain the two corresponding
cross sections ($\sigma_i$) and associated errors ($\delta_i$):
 \begin{eqnarray*}
 \sigma_i=\frac{S_i}{\mathcal{L}\;\;\epsilon_i\;\;b_i},
 {\,\,\,\,\,\,\,\,}
 \delta_i=\frac{\sqrt{S_i+B_i}}{\mathcal{L}\;\;\epsilon_i\;\;b_i}
 \end{eqnarray*}
with $b_i$ the corresponding branching ratios, $\mathcal{L}$ the integrated luminosity
and  $\epsilon_i$ the CLAS detection efficiency.

The cross section for a possible $\Theta^+$ is then built as the weighted average 
of  the two, using the CLAS efficiency  evaluated using  the five models 
described in Sec.\ref{ssec:ul}. The dependence on the model assumptions resulted to be within 30\%.
The largest cross section was obtained in the 
hypothesis of $t-$exchange dominance then chosen for conservatism.


\begin{table}[!h]
\caption{The $\Theta^+$ total cross section (nb) predicted by different models assuming  $\Gamma_{\Theta^+}=1$~MeV.}
\label{table:xs}
\begin{tabular}{| c | c | c | c | c |}
\hline
$J^P$         	& $1/2^+$ & $1/2^-$ & $3/2^+$ & $3/2^-$  \\
\hline\hline
\cite{oh04}    	& 100     & 0.4     &         &         \\
\cite{Guidal}	& 0.22    & 0.1    &         &             \\
\cite{Nam}   	& 2.0     &          & 1.0       & 3.0        \\
\cite{Roberts}	& 6.9     & 3.4     & 3.2     & 17.7    \\
\cite{Ko}     	& 15.0    &         &         &             \\
\hline
\end{tabular}
\end{table}

\begin{table}[!h]
\caption{Upper limits on the $\Theta^+$ width (MeV)  assuming a 95\% C.L. of 0.7 nb on the $\Theta^+$ total cross section.}
\label{table:gamma}
\begin{tabular}{ | c | c |  c | c | c |}
\hline

$J^P$         	& $1/2^+$  & $1/2^-$   & $3/2^+$ & $3/2^-$ \\
\hline\hline
\cite{oh04}    	& 0.01     & 1.8       &         &         \\
\cite{Guidal}	& 3.2      & 7.0       &         &         \\
\cite{Nam}   	& 0.35     &           & 0.7	 &  0.23    \\
\cite{Roberts}	& 0.10     & 0.2       & 0.2     &  0.04   \\
\cite{Ko}     	& 0.05     &           &         &         \\
\hline
\end{tabular}
\end{table}

\subsection{Upper limits on $\Gamma_{\Theta^{+}}$}
The $\Theta^{+}$ production cross section is directly connected to the
$\Theta^{+}$ width $\Gamma_{\Theta^{+}}$ (see for example Ref.~\cite{Roberts}).
Therefore upper limits on the  cross section imply  upper limits on the  resonance width. 
However, this connection  depends strongly on the theoretical model,
differing by  more than an order of magnitude for the available calculations~\cite{Roberts,Ko,oh04,Guidal,Nam}.
In Table~\ref{table:xs}, we summarize various theoretical predictions for the total
cross section averaged in the energy range 2--4 GeV for different assumptions for 
parity and spin of the $\Theta^+$ and  $\Gamma_{\Theta^+}=1$~MeV. 
For example,  assuming $J^P=1/2^+$ our  upper limit of 0.7~nb 
on  $\sigma ( \gamma p\to \bar K^0 \Theta^+)$
results in a  $\Gamma_{\Theta^+} < 3.2$\  MeV within the Regge approach
of Ref.~\cite{Guidal} and $\Gamma_{\Theta^+} < 0.35$ MeV for the other models.
The upper   bounds on the $\Theta^+$ width for these models are summarized in Table~\ref{table:gamma}.

\section{\label{sec:con} Conclusions}
In this paper we report the results of the first Jefferson Lab
high statistics and high resolution experiments entirely devoted to the pentaquark search on a nucleon target.
The reactions $\gamma p \to \bar K^0 K^+ n$ and $\gamma p \to \bar K^0 K^0 p$  were studied  in search of 
evidence of the $\Theta^+$ pentaquark in the $nK^+$ and $pK^0$ decay channels. 
The final states were isolated by detecting the 
$K^+$ or proton,  the  $K_S$ via its decay to $\pi^+\pi^-$
 and identifying the neutron or the second neutral kaon with the missing 
mass technique. 
For the former decay mode, the direct measurement of the $K^+$ allows the definition of the strangeness of any 
resonance observed in this final state.
A total of 160,000 and 550,000 events were selected for the reaction  $\gamma p \to \bar K^0 K^+ n$ and 
 $\gamma p \to \bar K^0 K^0 p$ respectively,  after 
the exclusion of background reactions.
The $\Theta^+$ was searched for as a narrow resonance in the $nK^+$ and $pK_S$ mass spectra
with a width of 3--4 MeV corresponding to the CLAS resolution for these channels in the kinematic region
of this experiment. Both mass spectra were found to be smooth and structureless.
No evidence for a narrow resonance was found in the mass range  1520--1600 MeV.
Combining the results of the two decay modes, we set an upper limit of 0.7 nb
(95\% confidence level) on the total production  cross section for the reaction $\gamma p \to \bar K^0 \Theta^+(1540)$.
This contradicts the results previously 
reported for a resonance in the reaction channel $\gamma p \to \bar K^0 K^+ n$.

\begin{figure}
\vspace{7.5cm} 
\includegraphics{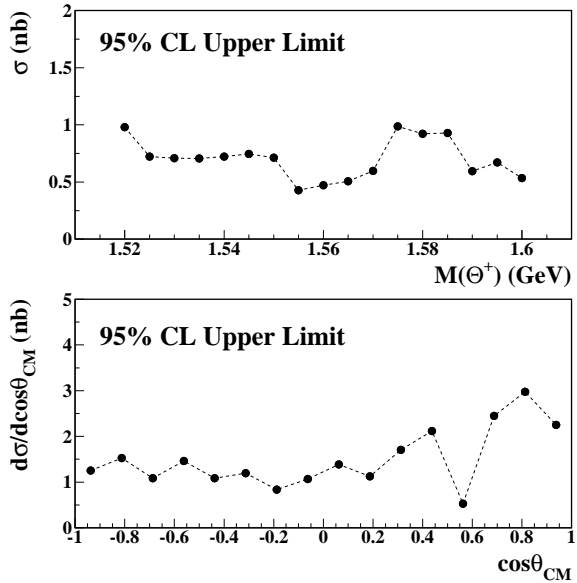}
\caption[]{The  95\% C.L. upper limit for the reaction  $\gamma p \to  \bar K^0 \Theta^+$
combining the two channels $\gamma p \to \bar K^0 K^+ n$
and $\gamma p \to \bar K^0 K^0 p$. 
The 95\% C.L. upper limit on the $\Theta^+$ cross section as a function of $\Theta^+$ mass (top) and
the 95\% C.L. upper limit on the  differential cross section $d\sigma/d\cos\theta^{CM}_{\bar K^0}$
for a fixed  $\Theta^+$ mass of 1540 MeV. The dashed line is to guide the eye.} 
\label{fig:final_xs}
\end{figure}

The accuracy in the mass determination was found to be 1--2 MeV from the comparison of 
the measured masses of known particles with  world data.
The quality of the data and the analysis 
procedures were tested by deriving the differential and the  total cross section for some known 
reactions and  obtaining an agreement within the experimental errors with existing measurements.
The same conclusions were found by several independent analyses, giving confidence in our final results.

Due to the loose hardware trigger of the experiment,
the same data set was analyzed to study the reactions 
$\gamma p \to \bar K^* \Theta^+$ and $\gamma p \to  K^- \Theta^{++}$ \cite{vpk++}. 
These findings, together with the results  coming from other pentaquark search experiments  at Jefferson Lab, 
could clear up the debate about the existence of the pentaquark.

\section{\label{sec:ack} Acknowledgments}
We would like to acknowledge the outstanding efforts of the staff of the Accelerator
and the Physics Divisions at Jefferson Lab that made this experiment possible. 
This work was supported in part by  the  Italian Istituto Nazionale di Fisica Nucleare, 
the French Centre National de la Recherche Scientifique
and Commissariat \`a l'Energie Atomique, 
the U.S. Department of Energy and National Science Foundation, 
and the Korea Science and Engineering Foundation.
The Southeastern Universities Research Association (SURA) operates the
Thomas Jefferson National Accelerator Facility for the United States
Department of Energy under contract DE-AC05-84ER40150.

\end{document}